\documentclass[acmsmall,nonacm,screen]{acmart}


\usepackage{booktabs}   
\usepackage{subcaption} 
\usepackage{xspace}
\usepackage{xparse}
\usepackage{microtype}
\usepackage{listings}
\usepackage{tikz}
\usepackage{float}
\usepackage{multicol}
\usepackage{amsbsy}
\usepackage{wrapfig}
\usepackage{mathtools}

\usetikzlibrary{calc}

\usepackage{subcaption}
\usepackage{mathtools}
\usepackage{amsbsy}
\usepackage{float}
\usepackage{microtype}
\usepackage{listings}
\usepackage{wrapfig}
\usepackage[group-separator={,},binary-units=true]{siunitx}

\newcommand{\todoinline}[1]{}

\newcommand{\ignore}[1]{}

\newcommand{\figref}[1]{\autoref{#1}}
\newcommand{\secref}[1]{\autoref{#1}}

\newcommand{\appendixPlaceholderImpl}{Appendix B}
\newcommand{\appendixNumbers}{Appendix C}

\setcopyright{acmcopyright}
\copyrightyear{2021}
\acmYear{2021}
\acmDOI{}

\acmConference[]{}{}{}
\acmBooktitle{Woodstock '18: ACM Symposium on Neural Gaze Detection,
  June 03--05, 2018, Woodstock, NY}
\acmPrice{15.00}
\acmISBN{978-1-4503-XXXX-X/18/06}

\definecolor{keywordcolor}{rgb}{0.5,0,0.5}
\definecolor{textgray}{gray}{0.4}
\definecolor{mygray}{rgb}{0.5,0.5,0.5}
\newcommand\code[1]{\lstinline[mathescape=true,basicstyle=\ttfamily\normalsize]|#1|}

\newcommand{\llvm}{LLVM}

\newcommand{\ozero}{\texttt{-O0}}
\newcommand{\oone}{\texttt{-O1}}
\newcommand{\otwo}{\texttt{-O2}}
\newcommand{\othree}{\texttt{-O3}}

\newcommand{\cp}{copy-and-patch}
\newcommand{\cpcap}{Copy-and-Patch}
\newcommand{\cpshort}{C\&P}
\newcommand{\X}{$\times$}

\newcommand{\stencilNum}{\num{98831}}
\newcommand{\stencilSize}{\SI{17.5}{\mega\byte}}

\newcommand{\wasmStencilNum}{\num{1666}}
\newcommand{\wasmStencilSize}{\SI{35}{\kilo\byte}}

\newcommand{\llvmSize}{\SI{22.8}{\mega\byte}}
\newcommand{\stencilGenTime}{\SI{14}{\minute}}
\newcommand{\stencilGenThreads}{6 threads}
\newcommand{\astNodeNum}{\num{25}}

\begin{document}

\title{Copy-and-Patch Compilation}
\subtitle{A fast compilation algorithm for high-level languages and bytecode}


\author{Haoran Xu}
\affiliation{%
    \institution{Stanford University}
    \streetaddress{353 Jane Stanford Way}
    \city{Stanford}
    \state{CA}
    \postcode{94305}
    \country{USA, haoranxu@stanford.edu}
}
 \email{haoranxu@stanford.edu}
 
\author{Fredrik Kjolstad}
\affiliation{%
    \institution{Stanford University}
    \streetaddress{353 Jane Stanford Way}
    \city{Stanford}
    \state{CA}
    \postcode{94305}
    \country{USA, kjolstad@cs.stanford.edu}
}
 \email{kjolstad@cs.stanford.edu}

\renewcommand{\shortauthors}{H. Xu and F. Kjolstad}

\begin{abstract}
    Fast compilation is important when compilation occurs at runtime, such as query compilers in modern database systems and WebAssembly virtual machines in modern browsers. We present copy-and-patch, an extremely fast compilation technique that also produces good quality code. It is capable of lowering both high-level languages and low-level bytecode programs to binary code, by stitching together code from a large library of binary implementation variants. We call these binary implementations stencils because they have holes where missing values must be inserted during code generation. We show how to construct a stencil library and describe the copy-and-patch algorithm that generates optimized binary code.
    
    We demonstrate two use cases of copy-and-patch: a compiler for a high-level C-like language intended for metaprogramming and a compiler for WebAssembly. Our high-level language compiler has negligible compilation cost: it produces code from an AST in less time than it takes to construct the AST. We have implemented an SQL database query compiler on top of this metaprogramming system and show that on TPC-H database benchmarks, copy-and-patch generates code two orders of magnitude faster than LLVM -O0 and three orders of magnitude faster than higher optimization levels. The generated code runs an order of magnitude faster than interpretation and $14\%$ faster than LLVM -O0. Our WebAssembly compiler generates code \num{4.9}\X{}--\num{6.5}\X{} faster than Liftoff, the WebAssembly baseline compiler in Google Chrome. The generated code also outperforms Liftoff's by $39\%$--$63\%$ on the Coremark and PolyBenchC WebAssembly benchmarks.
\end{abstract}


\maketitle

\section{Introduction}
\label{sec:introduction}

Fast compilation is important, particularly when the compilation occurs at runtime. Two representative use cases of runtime compilation are the WebAssembly virtual machines in modern browsers and the query engines of modern SQL databases, where WebAssembly modules and SQL queries are compiled to executable code and then executed. Since the latency experienced by the user is the sum of the time to generate the code (startup delay) and the time to execute the generated code, it is not enough to simply use the most optimizing but slowest compiler. The system must instead balance startup delay with execution performance. As an example, the TurboFan optimizing compiler~\cite{liftoffBlog} in the Google Chrome browser needs 49 CPU seconds to compile the WebAssembly module that powers the online AutoCAD Web App~\cite{autocadWebApp} on our system, which is too long for users to wait. As another example, the MemSQL~\cite{memsql} database takes up to 4.5 seconds to compile a TPC-H~\cite{tpch} database benchmark query using LLVM \othree{} on our system. And business-intelligence software may generate complex queries that take minutes to compile~\cite{memsqlPersonalCommunication}.

To balance startup delay and execution performance, runtime execution environment typically contain several execution tiers that occupy different points on the startup delay--execution performance Pareto frontier. Typical choices include interpreters, baseline compilers (also called template JITs), and choices of different optimizing levels for an optimizing compiler (e.g., LLVM \ozero{}, \oone{}, \otwo{}, and \othree{}). Modern databases, including Hyper~\cite{neumann2011}, Peloton~\cite{pelotonQueryCompilation}, PostgreSQL~\cite{postgresJit}, and MemSQL~\cite{memsql}, all employ tiered execution strategies that adds an interpreter, LLVM \ozero{}, or both below the most optimizing LLVM \othree{} tier.
Web browsers, on the other hand, use dedicated baseline compilers instead of interpretation or the \ozero{} version of an optimizing compiler. For example, the Google Chrome WebAssembly virtual machine first compiles using the fast Liftoff baseline compiler ~\cite{liftoffBlog} and then recompiles in the background using the optimizing TurboFan compiler. The motivation for using dedicated baseline compilers over \ozero{} compilation with optimizing compilers is startup delay. Baseline compilers such as Liftoff translate bytecode to machine code in one pass, without going through the intermediate representations of an optimizing compiler. As they move through the bytecode stream, they inspect each bytecode and emit appropriate machine code, either using a platform-dependent assembler as in Liftoff or pre-compiled from a high-level language like C or Java~\cite{maxine2013, templateJit2003, piumarta1998, ertl2003}. Thus, baseline compilers offer faster compilation than the \ozero{} compilation of an optimizing compiler---often measured in tens of megabytes of code generated per second---as well as better execution performance than an interpreter.

\begin{wrapfigure}{r}{0.4\linewidth}
    \includegraphics[width=\linewidth]{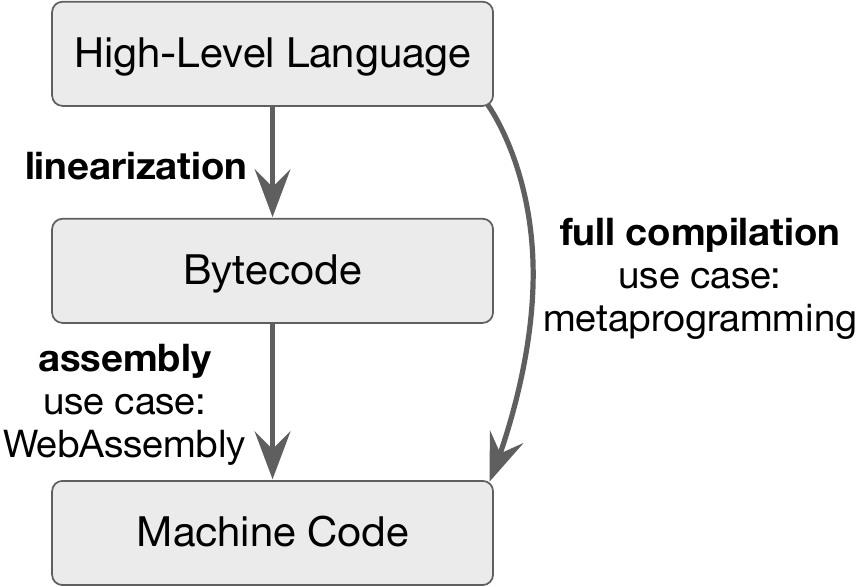}
    \caption{
        Phases of compilation: linearization, assembly, and full compilation.
        \label{fig:compilation-phases}
    }
\end{wrapfigure}

Interestingly, according to our survey of the 15 code-generating databases enumerated by  \citet{databaseOfDatabases}, not a single database has implemented a dedicated baseline compiler. This is not a coincidence. Database query compilers need to express the generated logic in a high-level metaprogramming language for expressiveness. And writing a baseline compiler to efficiently translate a high-level language with many complex language features directly to executable code is far more challenging than doing so for bytecode, where the input has already been processed into a stream of low-level opcodes that map closely to machine instruction. We therefore categorize compilers into \textit{full compilers} that compile from a high-level language to machine code and \textit{bytecode assemblers} that assemble a linearized stream of low-level bytecode to machine code, as illustrated in \figref{fig:compilation-phases}.

In this paper, we propose a new algorithm for template-JIT-style baseline compilers called \cp{}. Unlike prior baseline compiler techniques, \cp{} can be used to create both \textit{full compilers} and \textit{bytecode assemblers}. In addition, it shifts the startup delay--execution performance Pareto frontier by a large margin in both worlds. 

\begin{figure}
    \begin{minipage}[t]{0.48\textwidth}
        \centering
        \includegraphics[width=\textwidth]{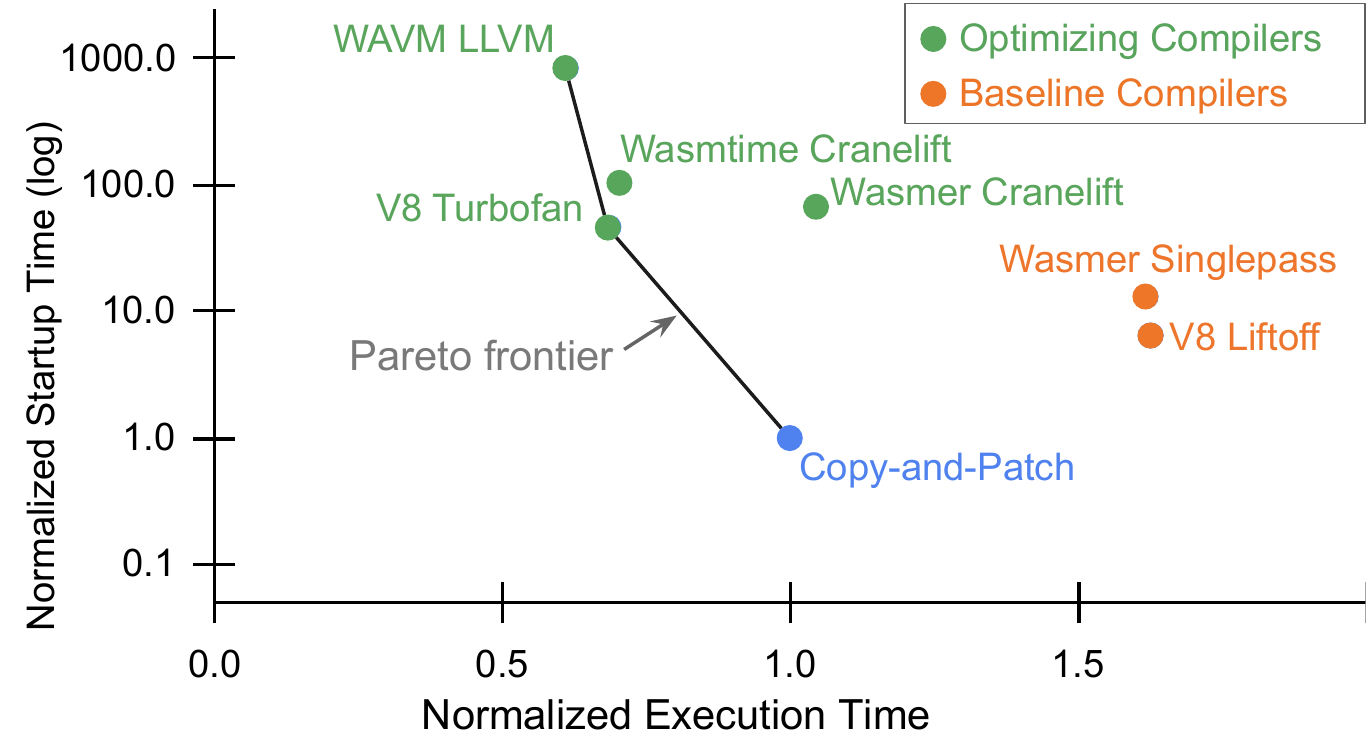}
        \caption{
            A scatter plot of normalized startup delay against execution time for seven WebAssembly compilers averaged over the PolyBench benchmarks. Our \cp{} compiler replaces baseline compilers on the Pareto frontier. 
            \label{fig:wasm-teaser}
        }
    \end{minipage}
    \hfill
    \begin{minipage}[t]{0.48\linewidth}
        \centering
        \includegraphics[width=\textwidth]{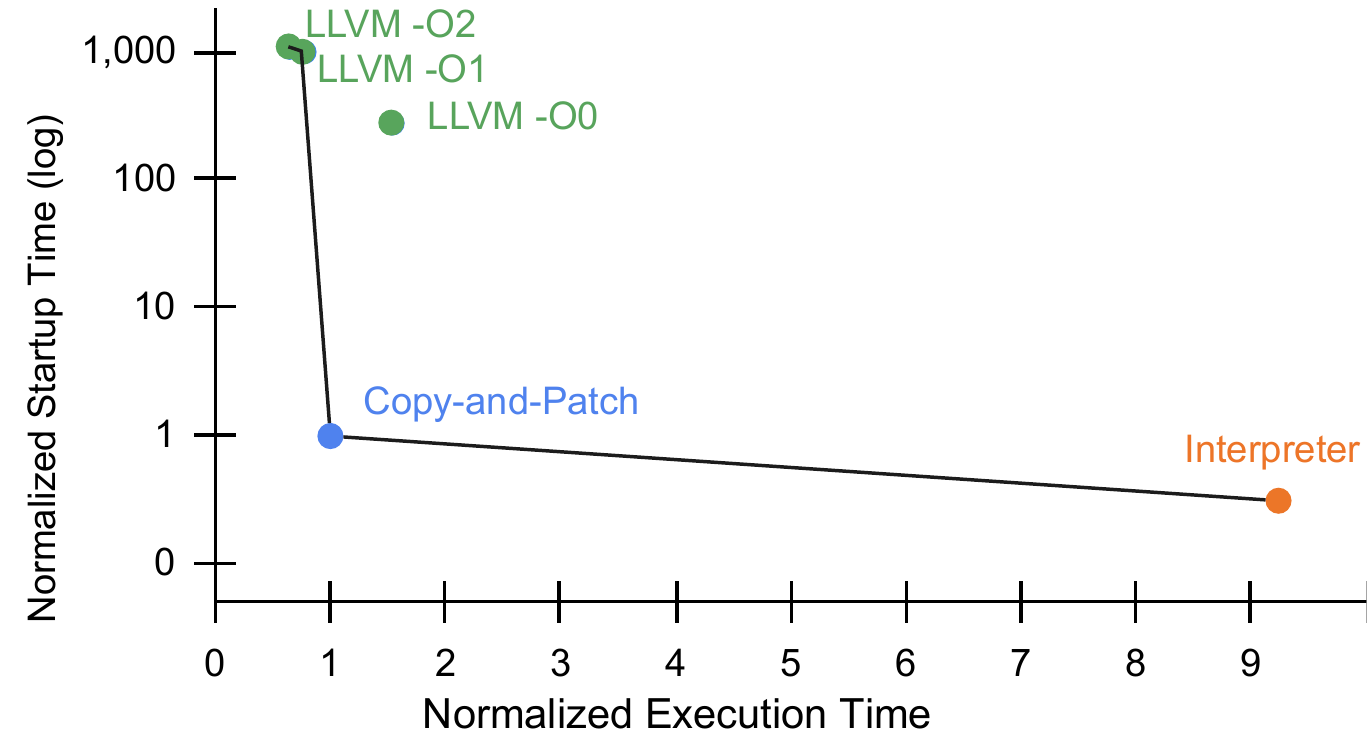}
        \caption{
            A scatter plot of normalized startup execution times against execution times for five strategies for executing the sixth TPC-H query implemented in a high-level language. Our \cp{} compiler replaces \llvm{}~\ozero{} on the Pareto frontier.
            \label{fig:hll-teaser}
        }
    \end{minipage}
\end{figure}

In the world of bytecode assemblers, we implemented a WebAssembly compiler based on \cp{}. Our compiler achieves both lower startup delay and better execution performance than prior baseline compilers. \figref{fig:wasm-teaser} shows the performance of six WebAssembly compilers on the PolyBenchC benchmark~\cite{polybenchc}, normalized to our performance. Our compiler has \num{6.5}\X{} lower startup delay than Liftoff, while generating on average $63\%$ better-performing code. In addition to Liftoff, our compiler also displaces Wasmer SinglePass~\cite{wasmerSinglepassJit}, the baseline compiler used in Wasmer~\cite{wasmer}, and even Wasmer Cranelift~\cite{wasmerCraneliftJit}, a relatively slow optimizing compiler, from the Pareto frontier.

In the world of full compilers, we used \cp{} to implement a compiler for a C-like high-level language intended for metaprogramming. The language is implemented as a DSL embedded in C++. To demonstrate that one can get a full compiler for a powerful language from \cp{} with reasonable efforts, our language supports a variety of complex features, including the C typesystem, all major C language constructs, the ability to use C++ classes and call C++ functions/methods in the host code, C++ exceptions semantics and destructor semantics, and so on. We then built a simple SQL database query compiler on top of this metaprogramming system. To the best of our knowledge, this implementation is the first database query compiler equipped with a dedicated baseline compiler. We evaluated its performance on eight TPC-H database queries~\cite{tpch}. \figref{fig:hll-teaser} shows the Pareto frontier of one of them. The compilation time of our compiler is so low that it is less than the time it takes to construct the AST of the program. Compared with interpreters, both have negligible startup delay (since constructing ASTs takes longer), but our execution performance is an order of magnitude faster. Compared with \llvm{}~\ozero{}, our implementation compiles two orders of magnitude faster and generates code that performs on average 14\% better. Therefore, we conclude that \cp{} renders both interpreters and \llvm{}~\ozero{} compilation obsolete in this use case.

At a high level, \cp{} works by having a pre-built library of composable and parametrizable binary code snippets that we call binary stencils.\footnote{The WebAssembly compiler uses \wasmStencilNum{} stencils taking \wasmStencilSize{} and the high-level compiler uses \stencilNum{} stencils taking \stencilSize{}.} At runtime, optimization and code generation become the simple task of looking up a data table to select the appropriate stencil, and instantiate it to the desired position by copying it and patching in the missing values. 
Our contributions are:
\begin{enumerate}
    
    \item The concept of a binary stencil, which is a pre-built implementation of an AST node or bytecode opcode with missing values (immediate literals, stack variable offsets, and branch and call targets) to be patched in at runtime.
    
    \item An algorithm that uses a library with many binary stencil variants to emit optimized machine code. There are two types of variants: one that enumerates different parameter configurations (whether they are literals, in different registers, or on the stack) and one that enumerates different code patterns (a single AST node/bytecode or a supernode of a common AST subtree/bytecode sequence).

    \item An algorithm that linearizes high-level language constructs like if-statements and loops, and generates machine code by composing multiple binary stencil fragments.
    
    \item A system called MetaVar for generating binary stencils, which allows the user to systematically generate the binary stencil variants in clean and pure C++, and leverages the Clang+LLVM compiler infrastructure to hide all platform-specific low-level detail.
    
\end{enumerate}

We evaluate our algorithm by evaluating the \cp{}-based compilers we built for WebAssembly and our high-level language.\footnote{Both compilers are open source under permissive licenses (MIT and Apache). The high-level language compiler is maintained at \url{https://github.com/sillycross/PochiVM} as part of a larger metaprogramming system project. The WebAssembly compiler is maintained at \url{https://github.com/sillycross/WasmNow}.} We compare the WebAssembly compiler to six industrial WebAssembly compilers, including those from Google Chrome and Wasmer. Our results show that our algorithm replaces all prior baseline compilers on the Pareto frontier and moves first-tier compilation closer to the performance of optimizing compilers. And we compare the high-level language compiler based on \cp{} with compiler implementations based on \llvm{} using different optimization levels, showing that our technique compiles two orders of magnitude faster than \llvm{}~\ozero{} while producing better code. We also show a breakdown of the performance contributed by different features in our compiler.

\section{Overview}
\label{sec:copy-and-patch-interpreter}


The topic of our paper is the \cp{} compilation algorithm and the associated MetaVar compiler. But to motivate their use and to evaluate them, we also built two compilers: one for WebAssembly and one for a high-level language. In this section, we first give an overview of the \cp{} algorithm and its surrounding ecosystem of tools and then give an overview of the two compilers.

\subsection{Copy-and-Patch and MetaVar Systems}

\begin{figure}
    \centering
    \includegraphics[width=0.58\linewidth]{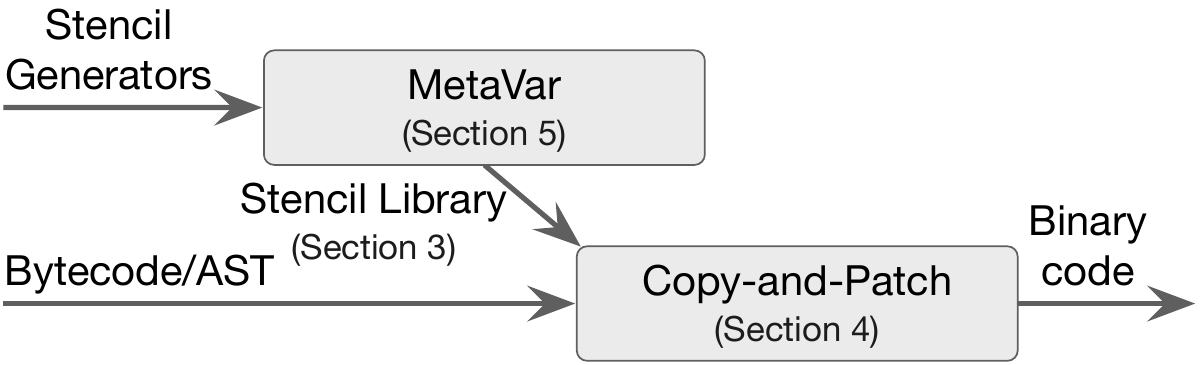}
    \caption{
        The copy-and-patch system compiles a high-level language AST or a bytecode sequence to binary code. It consists of the MetaVar compiler, which compiles stencil generators to binary stencils, and the \cpcap{} code generator, which generates the executable code for an AST or a bytecode sequence by copying and patching stencils.
        \label{fig:overview-system}
    }
    \vspace{-1em}
\end{figure}

The \cp{} system consists of two components: the MetaVar compiler and the \cp{} code generator. \figref{fig:overview-system} shows their relationship. The key to the \cp{} algorithm is the concept of a binary stencil, which is a partial binary implementation of a bytecode instruction or an AST node of a high-level language. The MetaVar compiler generates many binary stencils that implement different optimization cases for every bytecode or AST node. The MetaVar compiler takes as input bytecode/AST stencil generators and produces a library of binary stencils at library installation time. The stencil library becomes an input to the copy-and-patch code generator, together with a bytecode sequence or an  AST that implements a function. The code generator then produces binary code that implements the function, by copying and patching together stencils that implement the bytecodes or AST nodes. The patching step rewrites pre-determined places in the binary code, which are operands of machine instructions, including jump addresses and values of constants (stack offsets and literal values). Despite patching binary code, however, the system does not need any knowledge of platform-specific machine instruction encoding and is thus portable.

There are many stencil variants for each bytecode/AST node, and the copy-and-patch code generator produces optimized code by selecting among these variants. For example, if the AST contains an addition with a constant, then copy-and-patch will choose a variant that adds to a literal and then patch in the literal value. It will also perform register allocation by choosing among stencils that operate on values in registers and ones that operate on values on the stack, depending on register availability. And as a final example, the copy-and-patch algorithm will place the stencils of operators that follow each other in consecutive locations in memory, which lets it remove the jump between them. Together, these and other optimizations by stencil variant selection produces code that outperforms an interpreter, a bytecode baseline compiler, and even \llvm{} \ozero{}.

\subsection{Use case: WebAssembly Compiler}

WebAssembly is a bytecode format designed as a portable compilation target for programming languages, with the goal of enabling untrusted code to be executed safely and efficiently on any platform. Since WebAssembly code cannot be executed directly and an interpreter is too slow~\cite{liftoffBlog}, it must be assembled to machine code at runtime before it can execute. 

WebAssembly modules can often be large. For example, the AutoCAD Web App~\cite{autocadWebApp} is powered by a WebAssembly module of \SI{47.5}{\mega\byte}; and \textit{clang.wasm}~\cite{clangDotWasm}, the clang compiler in WebAssembly, is \SI{30.5}{\mega\byte}.  Since the user cannot interact with an application until the code starts executing, a fast baseline compiler is critical for a good user experience. As such, major web browsers like Chrome, Firefox, and Safari and major non-web WebAssembly runtimes like Wasmer and Wasmtime provide baseline compilers for WebAssembly, which prioritize a low startup delay at the expense of lower execution performance~\cite{liftoffBlog, firefoxWasmBaselineBlog1, firefoxWasmBaselineBlog2, safariWasmBlog, wasmerSinglepassJit, wasmtimeLightbeam}. On the other hand, performance is also important, since the major selling point of WebAssembly is that it lets native applications run at near-native speed on the Web, and the code generated by the baseline compiler will be executed until the optimizing compiler finishes, which can take a long time. The need for both extremely fast compilation and good execution performance makes WebAssembly a representative use case for \cp{}.

We implemented a WebAssembly baseline compiler using \cp{}. \figref{fig:overview-usecases-webassembly} shows its role in replacing the Tier 1 baseline compiler in a web browser, for both lower startup delay and better execution performance. Our compiler supports the full WebAssembly 1.0 core specification~\cite{wasmCoreSpec}, as well as a subset of the WASI embedding~\cite{wasmWasiEmbedding} necessary to run the benchmarks in \secref{sec:webassembly}. We note that an embedding only defines an agreement on how the imported functions of a WebAssembly module shall be implemented, so supporting more embeddings has nothing to do with code generation.

\subsection{Use case: High-Level Language Compiler}

\begin{figure}
    \begin{minipage}[t]{0.48\textwidth}
        \includegraphics[width=\linewidth]{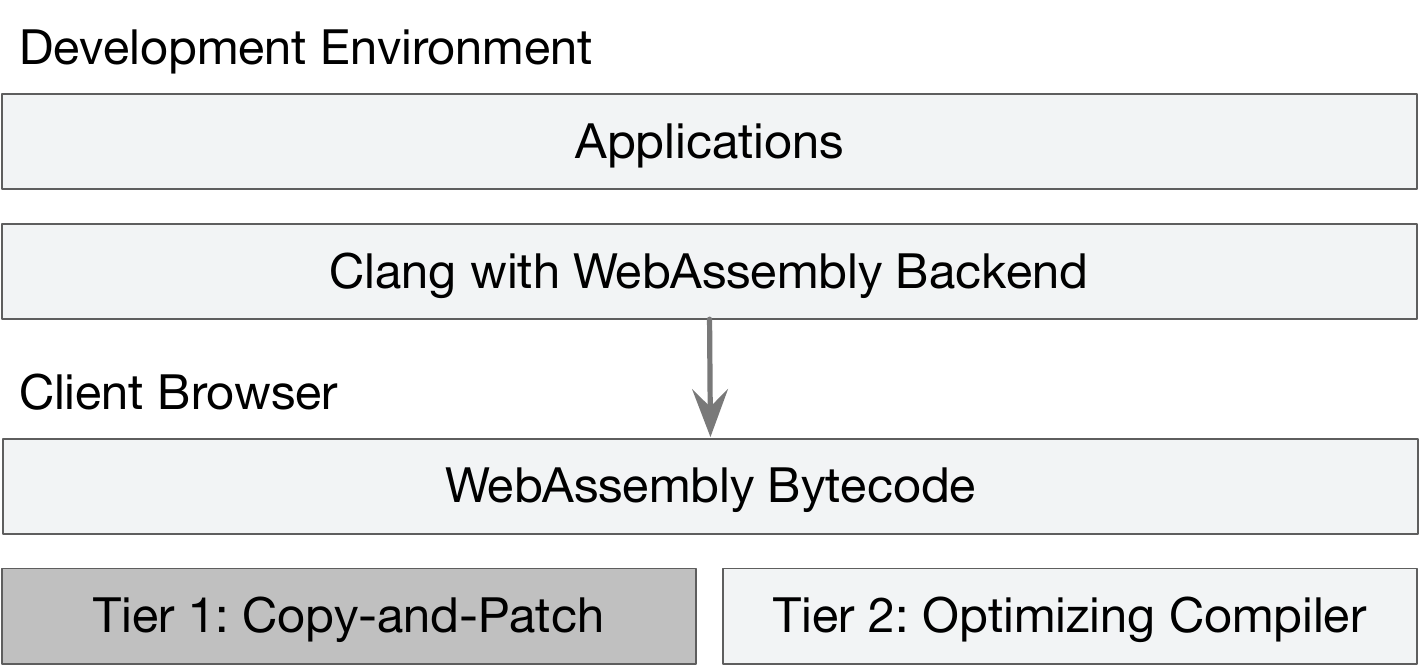}
        \caption{
             \cpcap{} can be used by WebAssembly compilers in browsers to both speed up Tier 1 compilation and produce better code. And it can be combined with a Tier 2 optimizing compiler to recompile hot code when the increased performance can amortize the orders of magnitude slower compilation.
            \label{fig:overview-usecases-webassembly}
        }        
    \end{minipage}
    \hfill
    \begin{minipage}[t]{0.48\textwidth}
        \includegraphics[width=\linewidth]{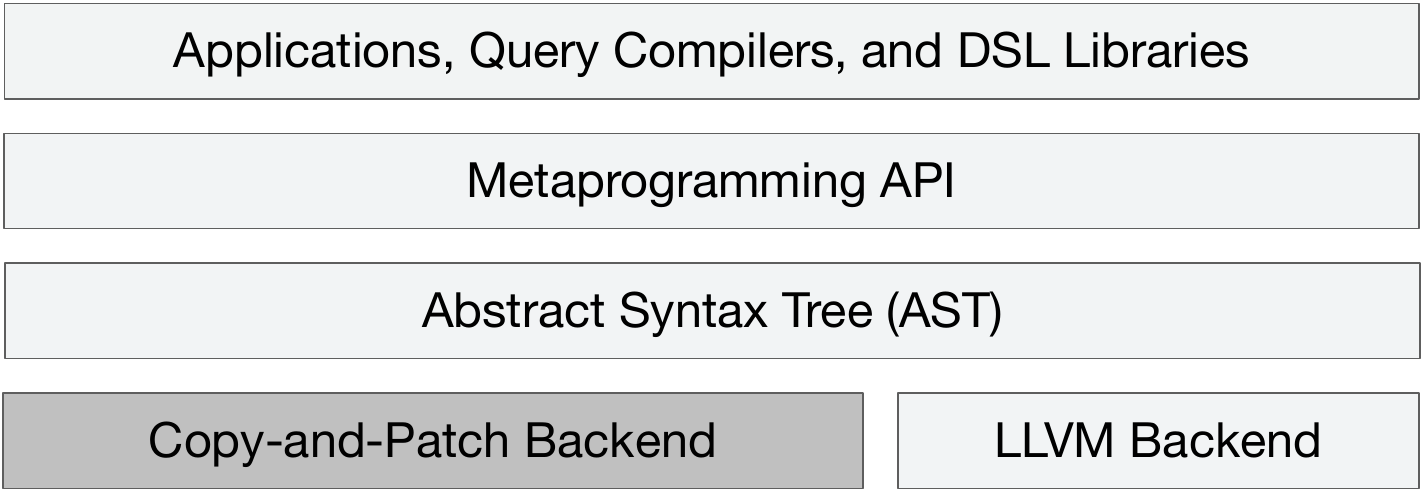}
        \caption{
            \cpcap{} can be used by metaprogramming systems to speed up their interpreters past LLVM \ozero{}. And it can be combined with LLVM to let the user request higher optimization levels when the increased performance can amortize the orders of magnitude higher compilation cost.
            \label{fig:overview-usecases-hll}
        }
    \end{minipage}
    \vspace{-1.4em}
\end{figure}

Our second use case for \cp{} is a metaprogramming language embedded in a C++ library that can be used to generate code at runtime. Example uses include database query engines and DSL libraries. Since a database query engine is expected to provide low latency, a metaprogramming system serving them must provide an execution path with low startup latency in addition to an optimizing path. In the database world the optimizing path is typically \llvm{} \othree{}, and the low startup latency path is either an interpreter or \llvm{} \ozero{}.
\figref{fig:overview-usecases-hll} shows the role of \cp{} in such a metaprogramming system. \cpcap{} provides better execution performance than \llvm{} \ozero{} at a negligible startup delay similar to an interpreter, thus rendering both interpreters and \llvm{} \ozero{} obsolete in this use case.
The user of this metaprogramming system can either use \cp{} or compile with LLVM \oone{} or higher at a much higher compilation time.

We have developed a compiler that implements the copy-and-patch code generation technique for the high-level metaprogramming language.
It supports all major imperative language constructs, local C++ objects, the ability to call external C++ functions, and C++ exceptions. Furthermore, users can expand the library with their own AST nodes, which is useful when a construct can not be implemented in the language and an external function call is too slow. For example, we provide an addition expression with C overflow semantics. But a user implementing a database query compiler might need an addition expression with SQL overflow semantics, implemented with low-level compiler intrinsics. Although the techniques we describe stand on their own, we believe our implementation can be used directly by metaprogramming systems such as Julia~\cite{bezanson2017}, Halide~\cite{ragan2012}, TACO~\cite{kjolstad2017}, Hyper~\cite{neumann2011}, Peloton~\cite{peloton}, and Terra~\cite{devito2013}.

\section{The Stencil Library}
\label{sec:stencil-library}
\label{sec:binary-stencils}

The stencil library contains binary implementations of bytecode or AST node types that are stitched together at runtime by the \cp{} algorithm to generate code for a bytecode module or a function in a high-level language. We call the binary implementations stencils, because they have holes where \cp{} inserts missing values to specialize them for the specific runtime AST. The stencil library contains many stencil variants for each bytecode or AST node type that are specialized for different operand types, value locations, and more. The variants let \cp{} optimize the generated code and do simple register allocation. Since the configuration options compose as a Cartesian product, the stencil library can grow to a significant number of stencils. Our WebAssembly implementation contains \wasmStencilNum{} stencils, taking \wasmStencilSize{} of memory. Our high-level language implementation is larger because it includes many supernodes and contains \stencilNum{} stencils, taking \stencilSize{} of memory. Although we believe these library sizes are practical for our respective use cases, there are too many stencils for it to be practical to write them by hand.

A binary stencil is a binary code function that implements a computation logic fragment, where literals, jump addresses, and stack offsets are missing. Each computation logic fragment implements the semantics of one AST node or bytecode, or a commonly-used shape of an AST subtree or bytecode sequence (we call such stencils supernodes). Supernodes allow optimizations across node boundaries, resulting in better quality of generated machine instructions. During \cp{} code generation, as we describe in \secref{sec:copy-and-patch}, the runtime AST or bytecode sequence are pattern-matched to stencils and supernode stencils. The selected stencils are copied and the missing values inserted. For example, if an instruction uses a literal then the value is filled in, and if it has a branch instruction to the next operation then the address is inserted.

\begin{wrapfigure}{r}{0.25\linewidth}
    \vspace{-.1em}
    \centering
    \hspace{-10mm}
    \vspace{-1em}
    \includegraphics[width=2.8cm]{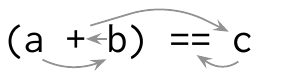}
    \hspace{-10mm}
    \caption{
        CPS
        \label{fig:continuation-passing}
    }
\end{wrapfigure}
The binary stencils use continuation-passing style (CPS)~\cite{steele1977} to pass control to the next stencil. With continuation-passing, control is passed directly to the next operation instead of being returned to the parent operation. \figref{fig:continuation-passing} shows how continuation-passing control flow moves bottom-up through an expression. Since function calls to pass on control are tail calls, the Clang C++ compiler that the MetaVar system uses to compile stencils lowers them to jump instructions. Combined with the GHC calling convention~\cite{ghcConvention}, in which all registers are saved by the caller and all parameters are passed in registers, continuation-passing removes most of the calling overhead between stencils.

Register allocation is another important optimization required for fast binary code. The obvious way to pass a temporary value between stencils is to reserve a slot in the stack frame whose offset is represented by a stencil hole. However, this is suboptimal in term of performance (since each read/write is a memory access), and we want to allow temporary values to be passed around in CPU registers. The trick to accomplish this also lies in the GHC calling convention, where all function parameters are passed in registers. Therefore, to pass a value as a parameter to the continuation is to pass this value in register to another stencil. In other words, we repurpose the function prototype and the calling convention as a register allocation protocol, where each function parameter implicitly corresponds to some physical register determined by the calling convention. We generate different variants of stencils with different function prototypes for different register configurations, so that at runtime we can pick the right one based on the circumstance.

\begin{wrapfigure}{r}{0.45\linewidth}
    \centering
    \includegraphics[width=\linewidth]{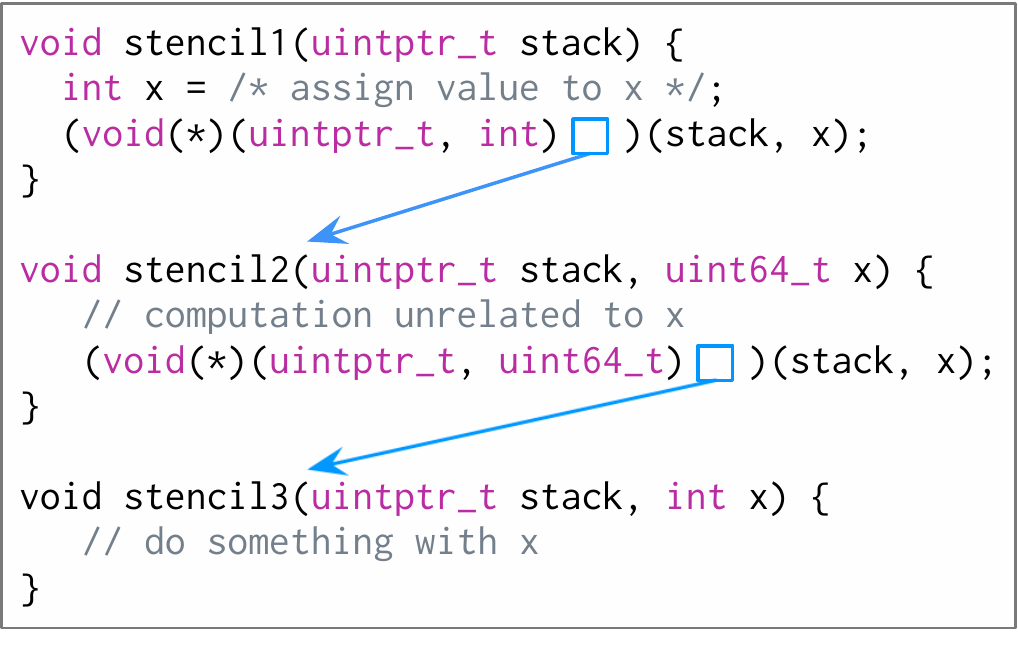}
    \caption{
        Three stencils that are executed in the order of the arrows. The first stencil produces a temporary value \code{x} that we want to pass in a register to the third stencil. The second stencil is executed in between, so it must not clobber the register. This is achieved by the pass-through parameter in the second stencil.
        \label{fig:passthrough-example}
    }
\end{wrapfigure}

We cannot naively enumerate all possible combinations of function prototypes for the different types of values that may be passed through, since the total number of combinations grows exponentially. The crucial observation is that each stencil only cares about its own inputs. The contents stored in the other registers do not matter, as long as they are not clobbered by the stencil. Therefore, for those registers, it is sufficient to always represent it by the longest type (\code{uint64_t} or \code{double}), and pass it from the argument to the continuation verbatim. We demonstrate this with a concrete example as shown in \figref{fig:passthrough-example}. In this example, we have three stencils. Stencil 1 produces a temporary value \code{x} of type \code{int}, which is to be consumed by stencil 3. But stencil 2 is executed in between, so it must be instructed to not clobber the register holding the value. As shown in the figure, stencil 1 calls its continuation with the temporary value \code{x} as a new parameter. This puts the value in the register. Stencil 2 does not care about what is stored in the register, but it must not clobber it. This is achieved by having \code{x} passed directly from the parameter to its continuation: we call it a pass-through parameter. There are two points worth mentioning. First, despite that the true type of \code{x} is \code{int}, the type of the pass-through parameter is \code{uint64_t}. This prevents the exponential explosion of different type combinations as explained earlier. Second, the value comes in as the second parameter and is passed to the continuation as the second parameter as well, which guarantees both correctness and performance. Correctness is guaranteed by the C language semantics: the callee should see whatever passed by the caller. Performance is guaranteed by the calling convention, due to which we can expect \code{x} to live in the same register. Therefore, passing \code{x} to the continuation is a no-op in the generated machine code. The pattern shown in the example can be generalized. A stencil can assign a value to a register by adding it as a new parameter to the continuation. A stencil can access the value by having a parameter with matching type at the same parameter ordinal, and can protect the value for future use by passing it through to the continuation. The lifetime of the value ends when it is no longer passed to the continuation, and the corresponding register is then free for future use.  In our current implementation, we only use registers to store temporary values while evaluating expression trees. However, we note that this mechanism can be used to implement the {\tt mem2reg} optimization to keep hot local variables in registers as well (see \secref{sec:copy-and-patch}).

\begin{wrapfigure}{r}{0.54\linewidth}
    \centering
    \includegraphics[width=\linewidth]{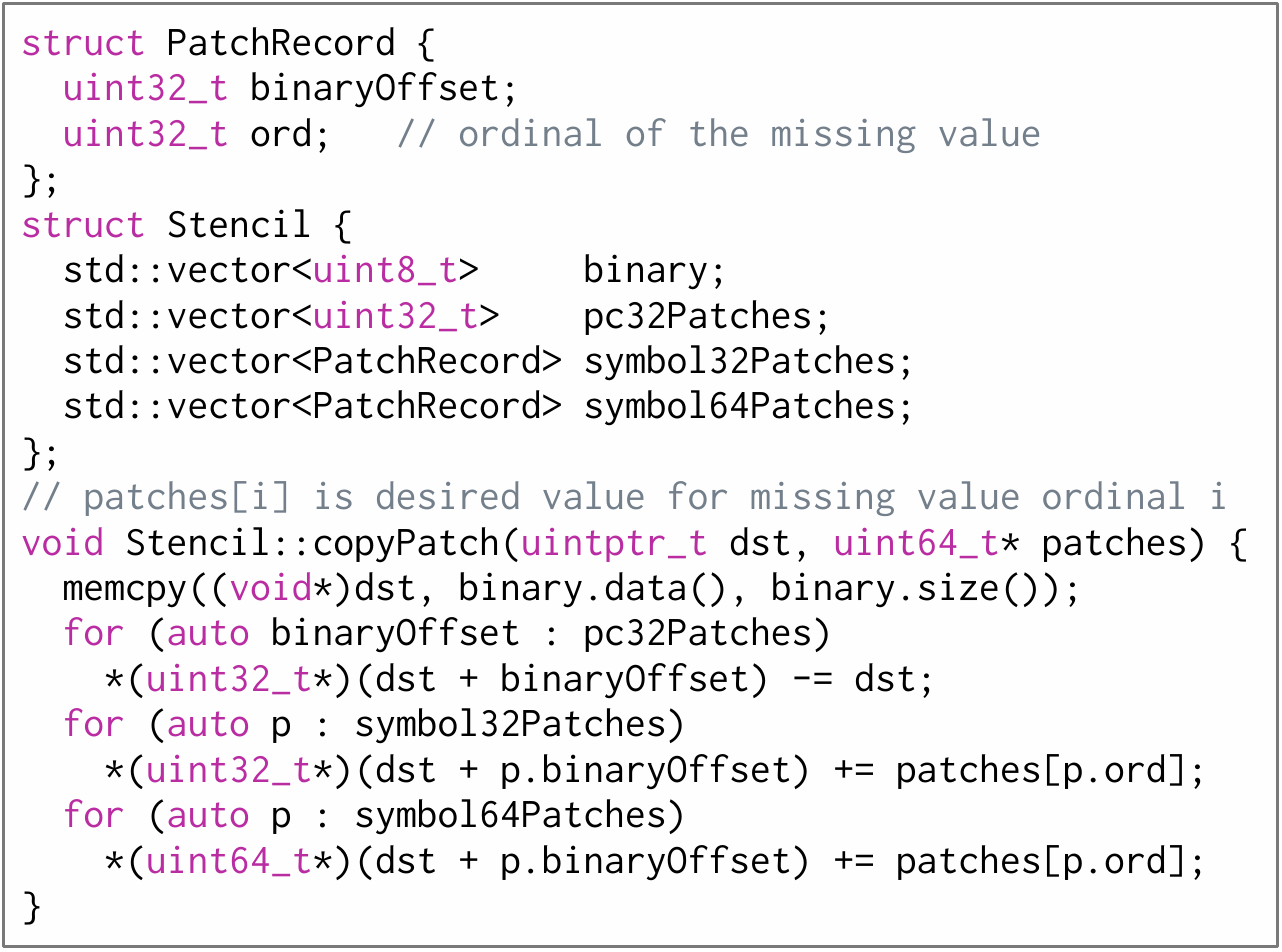}
    \caption{
        The data structure that stores a binary stencil and the method that materializes it at a given address. The struct includes the binary code and the locations of missing values to patch in during \cp{} code generation.
        \label{fig:binary-stencil-struct}
    }
\end{wrapfigure}

\begin{figure}
    \begin{minipage}[t]{0.49\textwidth}
        \includegraphics[width=\linewidth]{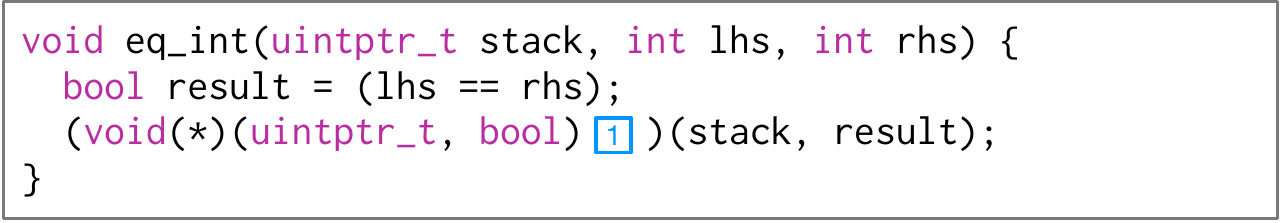}
        \subcaption{
            Equality test of two integer values in registers. Only the con\-tinuation-passing call to the next code fragment is missing.
            \label{fig:stencils-eq_int}
        }
    \end{minipage}
    \hfill
    \begin{minipage}[t]{0.49\textwidth}
        \includegraphics[width=\linewidth]{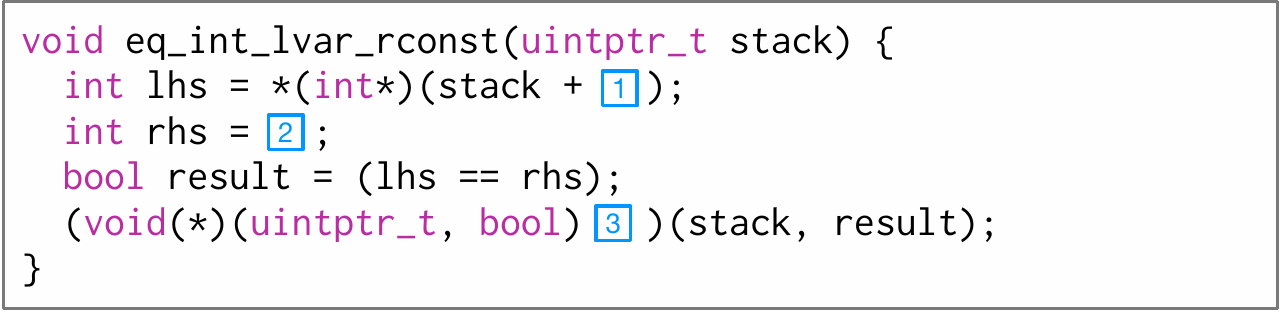}
        \subcaption{
            Equality test of an integer value on the stack to a constant. The first value's stack offset and the integer constant are missing.
            \label{fig:stencils-eq_int_with_right_const}
        }
    \end{minipage}
    
    \begin{minipage}[t]{0.49\textwidth}
        \includegraphics[width=\linewidth]{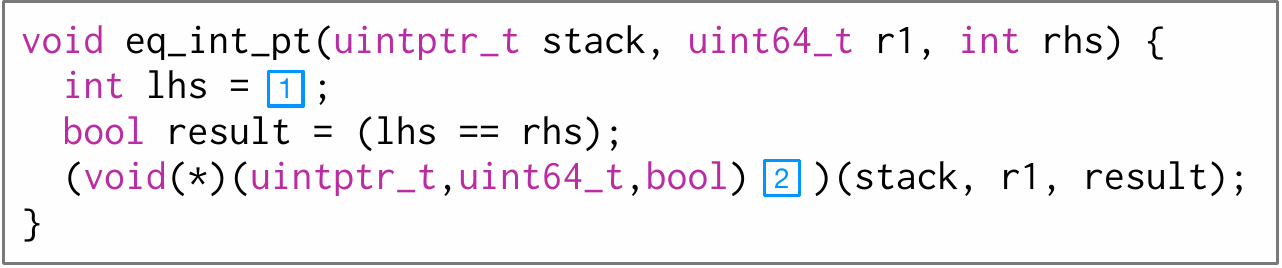}
        \subcaption{
            Equality that passes through a value stored in a register by a prior operation for use by a later operation.
            \label{fig:stencils-eq_int_pt}
        }
    \end{minipage}
    \hfill
    \vspace{-1em}
    \begin{minipage}[t]{0.49\textwidth}
        \includegraphics[width=\linewidth]{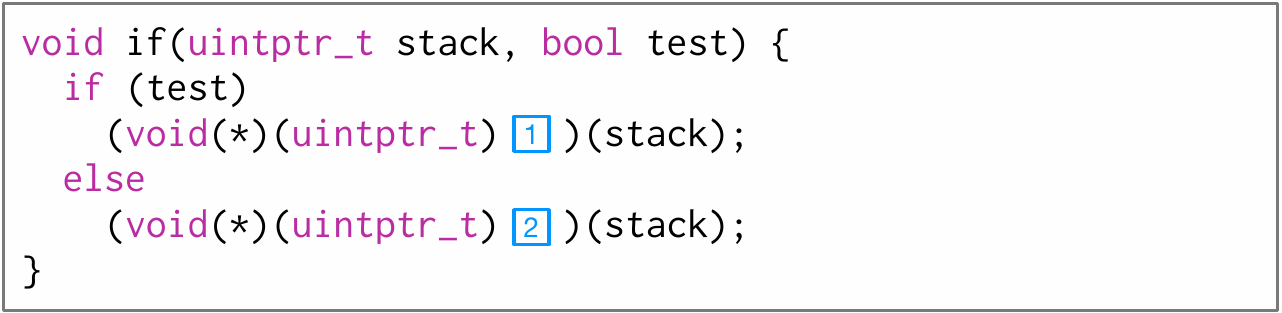}
        \subcaption{
            General if-then-else statement. The calls to the stencils of the then and else statements are missing.
            \label{fig:stencils-if}
        }
    \end{minipage}
    \caption{
        The conceptual logic of four out of \stencilNum{} stencils in the stencil library for the high-level language compiler, generated by instantiating 25 stencil generators written as C++ templated functions. Blue boxes indicate the holes in stencils, whose implementations are explained in \secref{sec:stencil-generators}. 
        \label{fig:stencil-example}
    }
\end{figure}

To summarize, the stencils come in many variants for each bytecode or AST node type, use continuation-passing style, and leave missing values to be filled in by the \cp{} algorithm. \figref{fig:stencil-example} shows conceptual C++ code for four stencils, replacing the missing values with blue numbered boxes. We show the implementations of these boxes in \figref{fig:generator}.  The C++ stencils are compiled by MetaVar at compiler installation time to produce binary stencils, as described in \secref{sec:metavar}. Figures~\ref{fig:stencils-eq_int}--\ref{fig:stencils-eq_int_pt} show three of the stencils that implement equality expressions. \figref{fig:stencils-eq_int} takes the expression's operands as arguments, compares them, and passes control and context to the next stencil by calling a continuation function. The blue box is the missing address to the next stencil. Missing addresses, and other types of missing values, are filled in when \cp{} generates code for an expression at runtime. Since MetaVar compiles the stencils with the GHC calling convention, the binary code assumes arguments are in registers. Thus, this stencil compares two integers whose values are stored in registers. \figref{fig:stencils-eq_int_with_right_const} is an equality variant where the first operand has been spilled to the stack, while the second operand is a literal. It has three missing values: the stack offset, the literal value, and the continuation address. And \figref{fig:stencils-if} shows the stencil for an if-then-else, taking the result of the test as an argument and calling one of two continuation functions.

The \cp{} algorithm composes stencils to implement an AST. It attempts to keep values in registers by selecting variants that work on arguments, like the one in \figref{fig:stencils-eq_int}. If a value is stored in a register and is needed by a later operation, then \cp{} chooses stencils variants that pass these values through, like the one in \figref{fig:stencils-eq_int_pt}. The pass-through stencil, by inserting arguments passed through to the continuation, forces Clang to ensure register values at function entry are unchanged when the continuation is called. This encourages Clang to use other available registers in the function body. When the available registers are insufficient to hold a temporary value for its lifetime, the \cp{} algorithm composes a variant that spills it to the stack with a variant that uses the spilled value.

The MetaVar system compiles C++ stencils to binary stencils that contain binary code and information about where the missing values are located. \figref{fig:binary-stencil-struct} shows the \code{Stencil} struct that stores a stencil. The \code{binaryCode} array contains the binary code, followed by three arrays that store the locations of the stencil's missing values, so that they can be patched in by the patching phase of \cp{}.

The stencil library maps stencil configurations that identify each stencil to the stencils themselves: $$(\text{configuration}) \rightarrow (\text{stencil}).$$ The configurations contain what AST node type the stencil implements, what types it operates on, whether it operates on constants, registers, or stack locations, and so forth.
The stencil library is generated by the MetaVar system at installation time, as described in \secref{sec:metavar}. The library is then used by the \cp{} algorithm in \secref{sec:copy-and-patch}, which weaves together the binary stencils for the AST nodes to create the generated function.

\section{Copy-and-Patch Code Generation}
\label{sec:copy-and-patch}

The \cp{} algorithm can be used to compile both bytecode and high-level languages. We will only describe compilation of high-level ASTs here, but the algorithm can be adapted to compile bytecode by removing the step that linearizes a high-level AST using the CPS graph.

The \cp{} binary code generation algorithm lowers an AST that describes a high-level language to binary code. It executes at runtime and produces code performing better than \llvm{} \texttt{-O0} at negligible cost. In most cases, compilation time is less than the time to construct the AST. The algorithm performs two post-order traversals of the AST: once to plan register usage, and once to select stencil configurations for AST nodes and construct a compact continuation-passing style (CPS) call graph. Next, the algorithm traverses the call graph depth-first. At each node, it copies the binary code of the node's stencil into the memory region immediately after the previously copied stencil. Finally, it patches the missing values into the stencil's binary code, including literal values used in the AST, stack offsets for local variables and spilled temporaries, and branch, jump, or call addresses to other stencils.

\begin{figure*}
    \begin{minipage}{\textwidth}
        \begin{minipage}[b]{0.45\linewidth}
            \centering
            \begin{minipage}{\linewidth}
                \centering
                \includegraphics[scale=0.8]{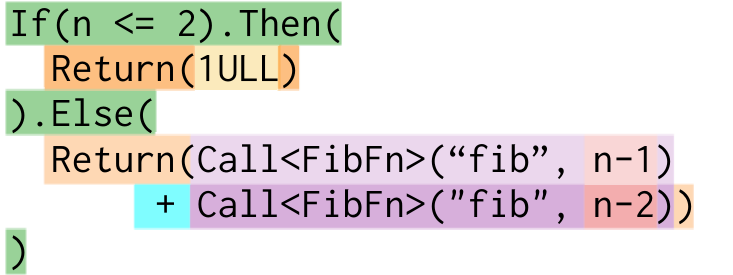}
                \subcaption{
                    C++ code that constructs the AST of the Fibonacci function.
                    \label{fig:fibonacci-example-ast}
                    \vspace{8.5mm}
                }
            \end{minipage}
            \begin{minipage}[b]{\linewidth}
                \centering
                \includegraphics[width=\linewidth]{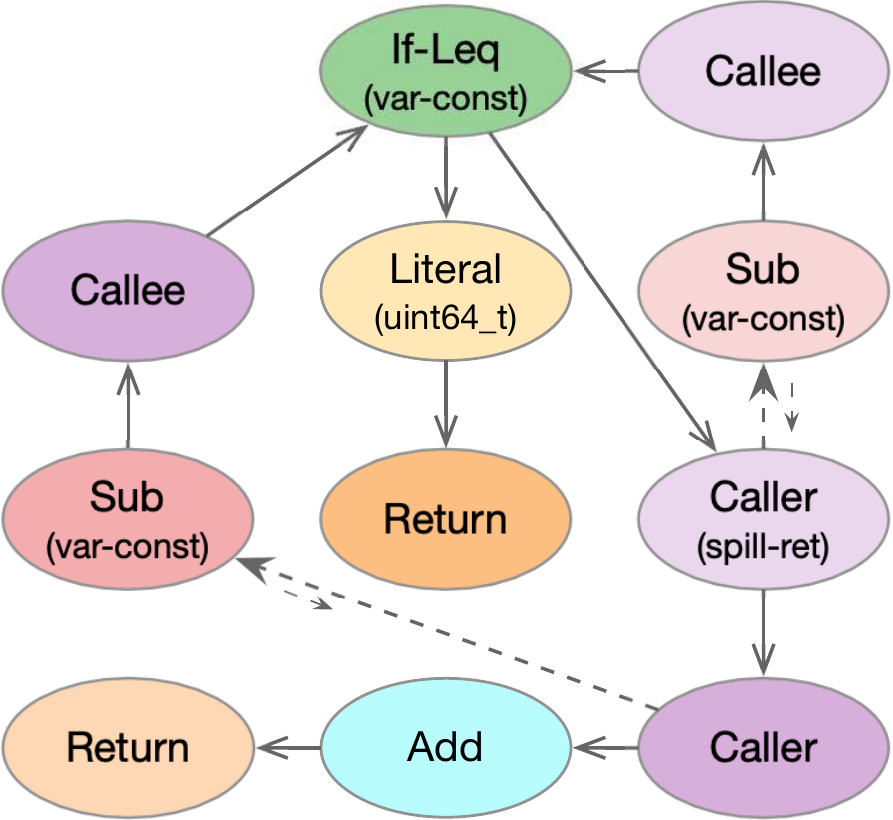}
                \subcaption{
                    CPS call graph between stencils. Solid arrows are tail calls compiled to jump instructions, while dotted lines require call instructions.
                    \label{fig:fibonacci-example-callgraph}
                    \vspace{-0.6em}
                }
            \end{minipage}    
        \end{minipage}
        \hfill
        \begin{minipage}[b]{0.51\linewidth}
            \centering
            \includegraphics[width=0.98\linewidth]{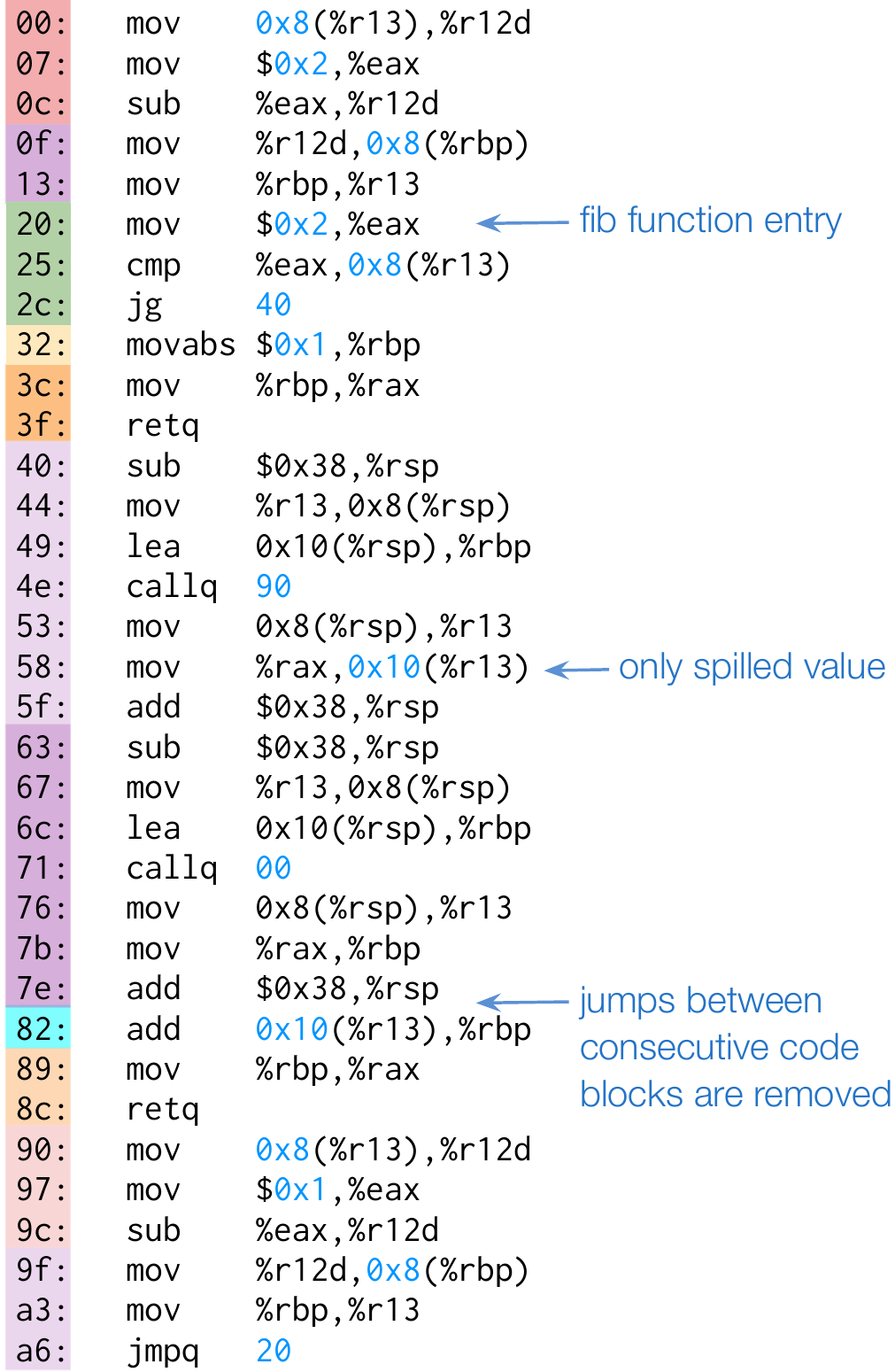}
            \subcaption{
                Assembly code generated by copying the call graph node stencils and patching in missing values (light blue).
                \label{fig:fibonacci-example-assembly}
                \vspace{-1.8em}
            }
        \end{minipage}
        \caption{
            Fibonacci function AST, its CPS call graph, and the assembly of the binary code generated by copying and patching each CPS call graph node. Each node or supernode has the same color in each representation.
            \label{fig:fibonacci-example}
            \vspace{1.5em}
        }
    \end{minipage}
    \begin{minipage}{\textwidth}
        \newcommand{\codescale}{0.6}
        \begin{minipage}{0.39\linewidth}
            \begin{minipage}[b]{\linewidth}
                \center
                \includegraphics[scale=\codescale]{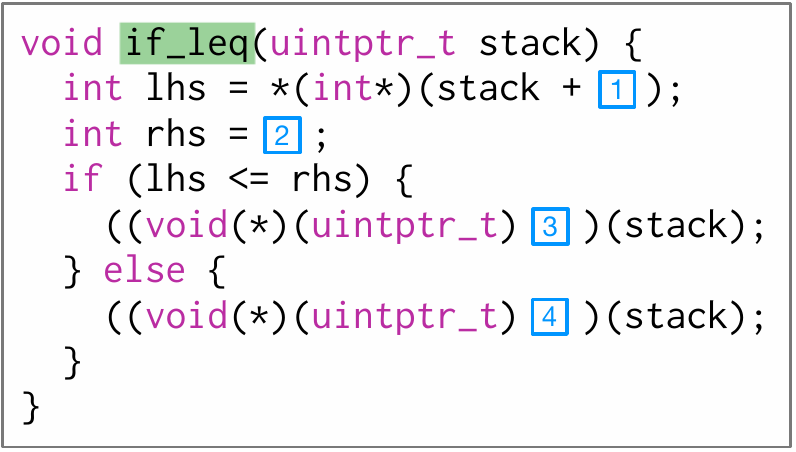}
                \subcaption{
                    The template-instantiated C++ logic for the If-Leq stencil in \figref{fig:fibonacci-example-callgraph}.
                    \label{fig:stencil-example-a}
                    \vspace{0.6em}
                }
            \end{minipage}
            
            \begin{minipage}[b]{\linewidth}
                \center
                \includegraphics[scale=\codescale]{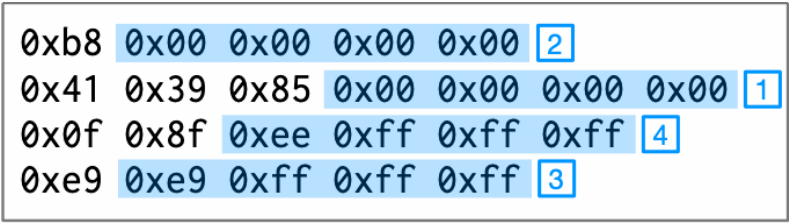}
                \subcaption{
                    Clang generates object code, with holes indicated by linker relocation records.\label{fig:stencil-example-b}
                    \vspace{-0.6em}
                }
            \end{minipage}
        \end{minipage}
        \hfill
        \begin{minipage}{0.59\linewidth}
            \begin{minipage}[b]{\linewidth}
                \center
                \includegraphics[scale=\codescale]{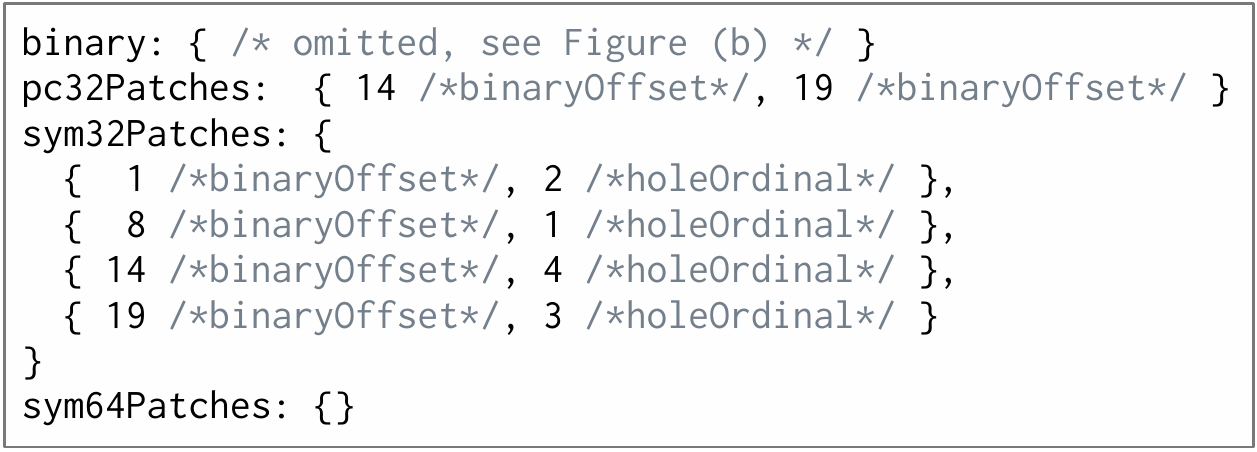}
                \subcaption{
                    The MetaVar compiler parses the object file, and generates the Stencil struct as described in \figref{fig:binary-stencil-struct}.
                    \label{fig:stencil-example-c}
                    \vspace{0.6em}
                }
            \end{minipage}
            
            \begin{minipage}[b]{\linewidth}
                \center
                \includegraphics[scale=\codescale]{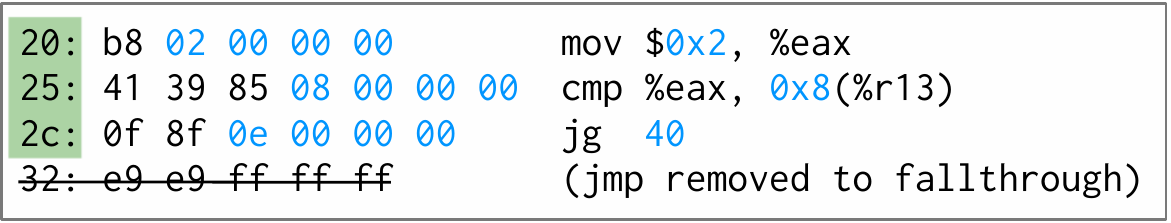}
                \subcaption{
                    At runtime, \cpshort{} generates executable code by copying the object code and patching the holes with runtime known values.
                    \label{fig:stencil-example-d}
                    \vspace{-0.6em}
                }
            \end{minipage}
        \end{minipage}
        \caption{
            The lifetime of the If-Leq (var-const) supernode stencil in \figref{fig:fibonacci-example-callgraph}.
            \label{fig:full-stencil-example}
        }
    \end{minipage}
\end{figure*}

An AST is lowered to binary code through a CPS call graph. \figref{fig:fibonacci-example} shows the representations that the Fibonacci function goes through on its way from the AST shown as printed code in \figref{fig:fibonacci-example-ast}, through the CPS call graph in \figref{fig:fibonacci-example-callgraph}, to machine code shown in assembly form in \figref{fig:fibonacci-example-assembly}. In each representation, the code that corresponds to each AST node is color-coded the same way, so that each node can be followed individually through the stages. We will use this example in our descriptions of each stage.

The \cp{} algorithm does light-weight register allocation to keep temporary values in registers to reduce the number of spills. In our benchmarked implementations, we only use registers to preserve temporary values produced while evaluating an expression. Specifically, given a budget on the maximum number of registers and an expression tree where each expression node produces a value, we perform register allocation using the algorithm described as follows. We perform a post-order traversal of the AST to abstractly evaluate the expression, and we always evaluate the children from left to right for each expression node. The traversal maintains the stack of outstanding temporary operands and, at each step, marks everything below the maximum number of register watermark to be spilled. Additionally, temporary values that cross a subsequent call must also be spilled because the GHC calling convention used for stencils assigns all registers to the callee (i.e., all registers are caller-saved). This algorithm is a simplified version of the Simple Sethi-Ullman Algorithm ~\cite{sethiUllman,sethiUllmanWiki} that does not choose between the orders of evaluating a node's children. We chose to use our modified algorithm primarily due to its very low overhead and little loss of practical effectiveness. The Simple Sethi-Ullman produces better decision than ours when it is beneficial to evaluate a right subtree first, but this requires the expression tree being both large and skewed toward the right side, which is not common. Nevertheless, we note that implementing the Simple Sethi-Ullman algorithm or the Advanced Seth-Ullman algorithm are both possible: it is only a trade-off between startup delay and execution performance, not a limitation of our technique.

As an example, \figref{fig:fibonacci-example-assembly} shows the effectiveness of our register allocation for the Fibonacci function: no temporary values were spilled except the result of the first function call on line 58, since its lifetime crosses the second function call. All other temporary values are passed in registers to the stencils that consume them: to name a few, the second Caller stencil to the Add stencil, the Add stencil to the Return stencil, and the Sub (var-const) stencil to the Callee stencil. There are also examples where a temporary value is protected by a pass-through parameter: the Caller stencil produces a temporary value of the new stack frame address, which is to be consumed by the Callee stencil. The Sub (var-const) stencil is executed in between, so a variant of the Sub (var-const) stencil that takes one pass-through parameter is selected to protect the value from being clobbered.

We have also explored the possibility of performing {\tt mem2reg} optimization to promote the storage of hot local variables from the stack to registers. Our prototype {\tt mem2reg} implementation gives up to 10\% execution performance boost, but results in about \num{3}\X{} slower compilation. We deemed this trade-off as not worthwhile in our use cases, so in our benchmarks, we choose to not perform this optimization. However, in other use cases it may be worthwhile to include the optimization.

During the AST traversal, we also plan the stack frame layout for the function, assigning a storage offset in the stack frame for each local variable and spilled temporary value. As another small optimization, once the lifetime of a spilled temporary value ends, its slot in the stack frame can be reused for another spilled temporary. This reduces stack frame size and improves locality.


The next step converts the AST to a CPS call graph, by selecting stencils that implement the AST nodes and linking them in the order they will be executed. The CPS call graph is constructed in the second post-order traversal of the AST. For each node, the algorithm selects the most specific stencil variant, depending on the AST tree shape and the context (e.g., whether the inputs live on the stack or in registers). It does a simple tree pattern matching to find sub-trees that can be implemented with an efficient supernode stencil. Supernodes allow Clang to optimize larger regions of code. For example, by leveraging advanced assembly instructions supported by the target architecture (e.g., advanced x86-64 addressing modes), an array access indexed by a constant or local variable can be compiled into fewer instructions. Having more supernodes allows better local optimizations and improves execution performance, at the cost of a slightly higher startup delay (since more branches are executed to perform the pattern matching) and a higher static memory footprint. Therefore, for our high-level language compiler, designed for database use cases in which memory footprint is less of a concern, we generated close to \num{100000} supernodes, covering a large set of common logic patterns in programs. One example is demonstrated by the If-Leq (var-const) supernode in \figref{fig:fibonacci-example-callgraph}, which implements an if-branch with a condition clause doing a less-or-equal comparison between a local variable and a constant \code{int}. We stress that this is not a special case, and not even the most complex case: just to name a few more examples, logic like \code{if (a[i] <op> b[j])}, or \code{c = a[i] <op> b[<literal>]} can be implemented by one supernode as well, for any compatible types of local variables \code{a,b,c,i,j}. Extensive supernode generation is made possible by our powerful MetaVar system, which allows us to systematically generate large numbers of supernodes easily (see \secref{sec: stencil-library-construction}). However, for applications where a small static memory footprint is desired (e.g., WebAssembly), we can simply remove the supernode stencils to get a small stencil library. As an example, the stencil library of our WebAssembly compiler is only \wasmStencilSize{}. The user can also add new supernode stencils specific to his/her use case, such as fused multiply-add. When a stencil is selected for one or several AST nodes, the stencil's configuration is added to the CPS call graph and a call edge is set to point to the next node in the post-order traversal. \figref{fig:fibonacci-example-callgraph} shows the call graph for Fibonacci. Most calls are tail calls (solid arrows), which the stencil compiler turns into jump instructions. Only two true calls are left: the two call expressions in the AST. 

The CPS call graph is then lowered to binary code by copying the binary stencil code of each node to contiguous memory. The copy step traverses the CPS call graph in depth-first order starting at any node that has no predecessors. At each call graph node, it retrieves the stencil corresponding to the node's configuration from the stencil library. It then copies the stencil's binary code into the memory region following immediately the binary code of the stencil belonging to the preceding call node. The purpose of copying these into consecutive locations in depth-first order is to maximize the number of stencil binary codes that jump to a stencil binary code right after it. As these jumps are fruitless, the stencil copy simply elides them from the copy. \figref{fig:fibonacci-example-assembly} shows the binary code in assembly form resulting from the Fibonacci function. Most jumps were successfully elided (e.g., line 82), while only two jumps could not be removed (line 2c and a6). If a stencil node has a fixed predecessor, and the predecessor is not a conditional branch, then our algorithm is guaranteed to elide the jump instruction. Therefore, all remaining jump instructions must correspond to some form of control-flow redirection statement (e.g., if-branches, loops, calls) in the input AST, and are thus necessary. 

The final step patches missing values into the copied binary code. For each stencil, it iterates through missing values to insert literal values from the AST, stack offsets for variables and temporaries, and branch, jump, or call targets to other stencils (for jumps that were not elided). \figref{fig:binary-stencil-struct} contains the stencil struct that stores information about missing values, and the logic to patch them. \figref{fig:fibonacci-example-assembly} shows the filled in missing values in blue. For instance, the jump target on line a6 jumps to the If-Leq (var-const) stencil on line 20, and the value 0x2 on line 20 is the literal 2 from the \code{int} constant literal AST node in the if-branch condition of \figref{fig:fibonacci-example-ast}.

The \cp{} technique also supports external function calls. External functions are important in database applications like the TPC-H queries in \secref{sec:database-benchmarks}, where data is stored in C++ data structures that must be iterated and accessed from generated code. The external call node expects the callee to take a single \code{void*} parameter, pointing to an array with the actual parameters. Template metaprogramming techniques can be used to automatically wrap any C++ function into this form, including functions with non-primitive parameters passed by value, overloaded functions, and method calls. In fact, the metaprogramming system we built on top of \cp{} supports calling \textit{any} C++ function in a type-safe manner. Code generated by \cp{} can also propagate C++ exceptions thrown by functions it calls. The wrapper function catches any thrown exception and stores it to a thread-local variable. The \cp{} external call node has a boolean return value that the wrapper functions use to signal that an exception was thrown. The return value is checked by generated code, and if true, branches to code that calls destructors, propagates it through the generated function call stack, and returns it to the calling host code that must re-throw the exception. Through these features, the copy-and-patch-generated code efficiently interoperates with the host language such as C++, allowing it to call host code and to manage the host code's exceptions. The description of a complete metaprogramming system built upon the \cpshort{} technique is, however, outside the scope of this paper.

\section{Stencil Library Construction}
\label{sec: stencil-library-construction}

The MetaVar compiler constructs the stencil library from programmer-specified stencil generators. The programmer specifies one stencil generator for each AST node by writing C++ code that uses template meta-variables to express variants, and uses special macros to express missing values to be patched at runtime. The MetaVar compiler then iterates at compile-time over the values of the meta-variables and instantiates the template for every valid combination. The instantiated templates are compiled by the Clang C++ compiler to object code. The stencil library builder then parses the object code to retrieve the stencil configurations and binary stencils, which are used to build the stencil library that is linked to the \cp{} runtime.

\subsection{Stencil Generators}
\label{sec:stencil-generators}

Stencil generators are templated C++ functions whose template instantiations produce stencils. Their template parameters are called metavars, which are defined by a fixed set of values. For instance, the \code{PrimitiveType} metavar enumerates primitive types, while a boolean metavar enumerates false and true. The MetaVar compiler iterates through the values of the metavars, subject to user-defined filter template functions, to instantiate stencils at library installation time.

\begin{wrapfigure}{r}{0.58\linewidth}
    \centering
    \includegraphics[width=\linewidth]{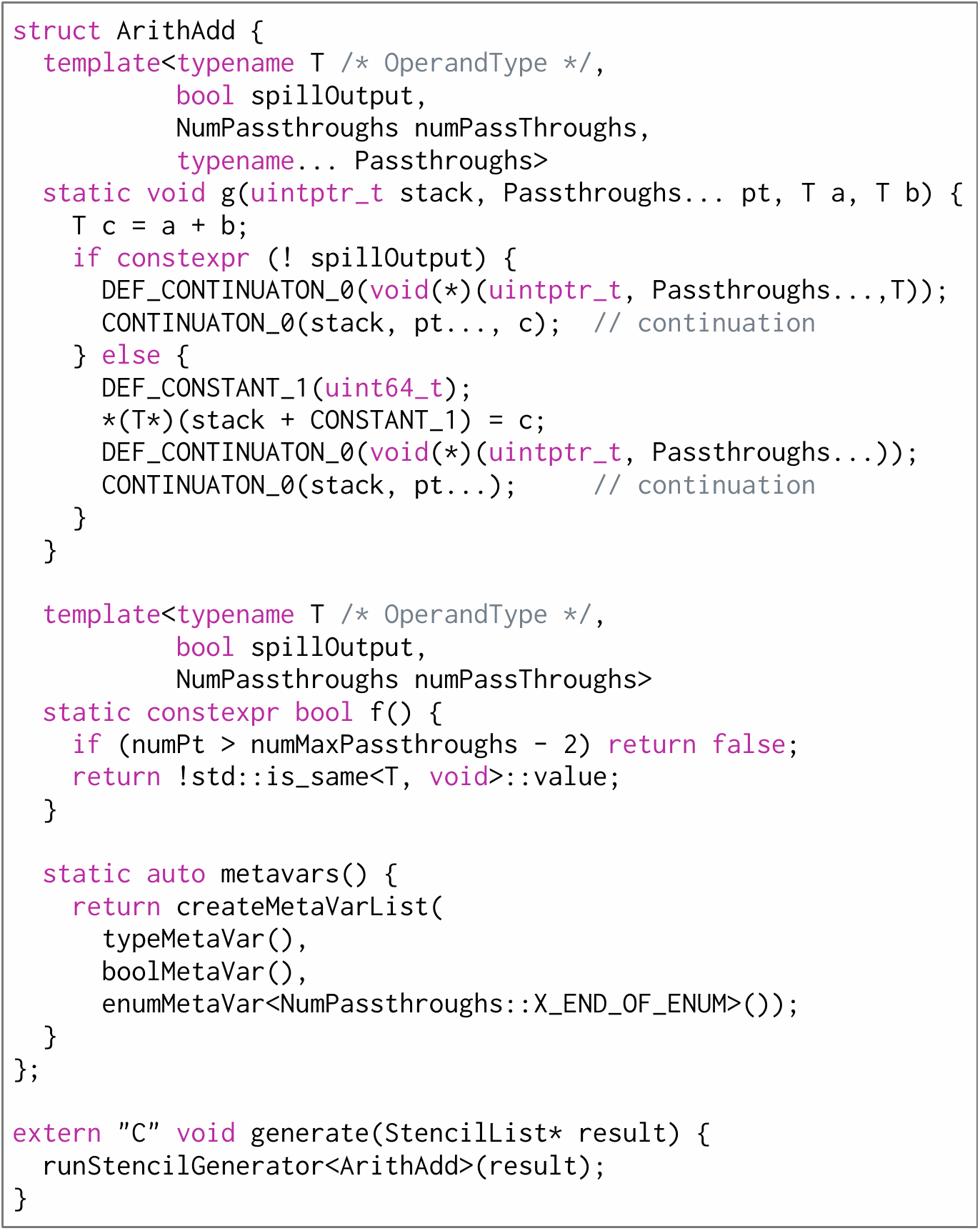}
    \caption{
        A simplified addition stencil generator written with C++ template metaprogramming. The generator is processed by MetaVar to generate stencils for addition.
        \label{fig:generator}
    }
\end{wrapfigure}

More precisely, let $S_1, \dots, S_n$ be the sets that enumerate the values of each metavar used in a stencil generator, where each $S_i$ is either a finite set of types or a finite set of values. Let $L = S_1 \times \dots \times S_n$ be their Cartesian product. Given a template filter function, $f : L \rightarrow \text{bool}$, and a template generator function $g : L \rightarrow \text{stencil}$, the MetaVar system generates a list of pairs $\langle l,g(l) \rangle$, for all $l \in L$ where $f(l)$ is true. In other words, the MetaVar system generates a list of tuples that map valid metavar configurations to C++ stencils.

Metavars and filters work together to produce valid stencils. Within stencils, missing functions and values are defined by special macros such as \code{DEF_CONTINUATION_0} and \code{DEF_CONSTANT_1}. We demonstrate these features by the simplified addition generator and filter functions in \figref{fig:generator}. The generator \code{g} has three metavars: the operand type \code{T}, whether the result should be spilled, and the number of pass-through variables to be preserved in registers across this stencil. It produces only stencils whose operands \code{a} and \code{b} are stored in registers, while a more sophisticated generator would support these having been spilled by a previous operation. It would also handle the case where one side has a simple shape (e.g. literal, variable, or simple array indexing). The compile-time conditional inside \code{g} determines whether the generated stencil spills the result or keeps it in a register. If spill is false, then the generated stencil simply passes the result \code{c} to the continuation as an argument, which will be stored in a register in the GHC calling convention. But if spill is true, then the result \code{c} is instead stored to the stack.

The missing values in a stencil are defined using special macros that also assign an ordinal to each missing value. For example, \code{DEF_CONTINUATION_0} defines ordinal 0 to be a function of a specified type, and \code{DEF_CONSTANT_1(int)} defines ordinal 1 to be a constant of type \code{int}. At runtime, the stencil can be patched by specifying the desired value for each ordinal, as shown in \figref{fig:binary-stencil-struct}. Internally, a special macro expands to a piece of code that declares a local variable of the given function or constant type and assigns to it the address of a pre-defined extern variable. We forcefully cast extern variables to their assigned types, using {\tt reinterpret\_cast} for functions and a C {\tt union} hack to bit-cast constants: see \appendixPlaceholderImpl{} in the supplemental material for the implementation. Since extern variables in C++ are by definition defined outside the current module, this forces Clang to emit information into the object code that identifies the locations of those missing values in the binary code. This information can be used to figure out how a stencil shall be patched, as elaborated in \secref{sec:metavar}.

The macro-defined constants can be used just like normal constant variables, except that you cannot compare equality between two macro-defined constants, or between such a constant and 0. The reason for these limitations is that two different extern symbols are never equal, and that symbols are never null. Additionally, in x86-64 architecture, MetaVar must compile the stencils using the suitable code model\footnote{There is an architecture idiosyncrasy involved here. To make the most efficient use of the x86-64 instruction set, Clang by default assumes that the address of an extern symbol fits in signed 32 bits (the ``small code model'' assumption), despite that the address is 64 bits. If the value we want to encode could be larger, we need to tell Clang the correct assumption.}~\cite{systemvABI}. Nevertheless, these limitations are easy to work around.

Finally, the pass-through arguments let the MetaVar system generate variants that avoid clobbering registers that the \cp{} algorithm uses for the results of other operations. For example, when computing $\text{term}_1 + \text{term}_2$, the result of the first term must be stored while computing the second term. In this case, the \cp{} algorithm would choose a stencil variant with one pass-through variable to protect the register that stores this result. The example in \figref{fig:generator} shows how pass-through template variables are used in a generator: they are passed from the arguments of the generator to its continuations. For ease of exposition, we only show one set of pass-through variables. In x86-64 integer and floating-point values are passed in separate sets of registers, so our implementation needs two sets of pass-through variables for integral and floating-point values respectively.

Any number of temporaries may need to be stored to compute an expression, but a given machine only has a limited number of registers. The filter function \code{f}, in addition to removing the \code{void} type that cannot be added, also bounds the number of pass-through variables for which stencils will be generated. As we have seen in \secref{sec:copy-and-patch}, the \cp{} algorithm chooses stencils to store temporary values in registers to avoid as many spills as possible, and uses pass-through stencils for subsequent operations. If a temporary cannot be kept in register for its lifetime, \cpshort{} will instead select the stencil that spill it to the stack.

\subsection{The MetaVar Compiler}
\label{sec:metavar}

\begin{figure*}
    \centering
    \includegraphics[width=\linewidth]{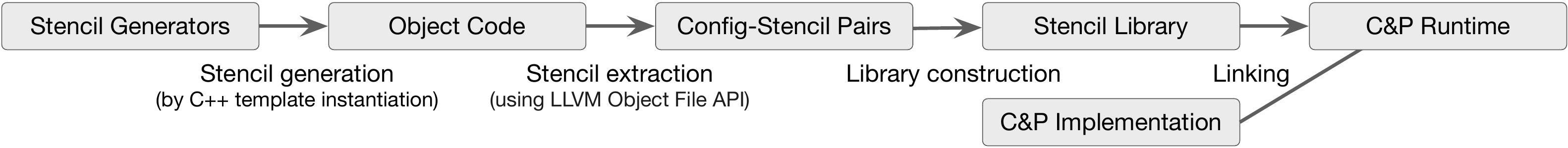}
    \vspace{-2em}
    \caption{
        MetaVar compiles stencil generators to a library that is linked to the \cp{} implementation.
        \label{fig:stencil-construction}
    }
    \vspace{-1em}
\end{figure*}

MetaVar compiles stencil generators to the stencil library, which maps stencil configurations to stencils. As described in \secref{sec:binary-stencils}, a stencil configuration describes a stencil, including what node type it implements and whether it spills the result, while the stencil consists of binary code and the locations of missing values. \figref{fig:stencil-construction} shows the stages of the MetaVar system as arrows, with boxes showing inputs and outputs. MetaVar leverages both C++ template metaprogramming and the Clang+LLVM compiler infrastructure to generate the binary stencils, avoiding the need to implement an optimizing compiler. The result is a concise system that can generate code for any platform that LLVM supports.

The first stage of the MetaVar system, stencil generation, converts stencil generators to stencils. Stencil generation is a C++ template program (the \code{runStencilGenerator} function in  \figref{fig:generator}) that for each stencil generator iterates over the Cartesian combination of metavar values using template recursion. For each combination of metavar values, which we call a configuration, the stencil generation checks its validity by calling the stencil generator's filter template function. It then stores all valid $\langle\text{configuration}, \text{function pointer}\rangle$ pairs into a list. As a side effect of taking the function pointer, all valid stencils are also instantiated by Clang and their implementations are compiled to object code.

The stencil extraction stage extracts configurations and corresponding stencils from the object code. The configurations are extracted by executing object code. As we saw in \figref{fig:generator}, a stencil generator contains a \code{generate()} stub function, containing a piece of boilerplate code that invokes our \code{runStencilGenerator} function and returns a list of stencil configuration and function pointer pairs. Stencil extraction uses the LLVM JIT machinery to execute this \code{generate()} function to retrieve the configuration. We then use an LLVM JIT API to get symbol names of the stencil functions from the function pointers in the list. The next step uses the LLVM object file parser to locate the functions based on the symbol names of the stencils we got in the previous step. It then extracts their binary code and the linker relocation records containing information about the extern symbols that were inserted by the placeholder macros, which is used to record the offsets to the missing values and how to patch them. Specifically, each missing value is a 32 or 64-bit scalar, and the patch shall be computed by initializing it with a fixed constant, optionally subtracting its memory address, and optionally adding the memory address of a symbol~\cite{systemvABI}. Although this computation rule is technically architecture-dependent, it applies to most major architectures including x86-64~\cite{systemvABI}, ARM~\cite{armRelocationTypes}, and SPARC~\cite{sparcRelocationTypes}. The rule yields the 3 patch vectors and the patch algorithm in \figref{fig:binary-stencil-struct} and, together with the binary code, give us a stencil.

The final stage constructs the stencil library from the configuration-stencil pairs. The library is a C++ file containing a static constant hash map that maps stencil configurations to the binary stencils in \figref{fig:binary-stencil-struct}, and the API for the runtime to select stencils from the hash map. The library is linked together with the \cp{} implementation to form the \cp{} runtime.

As a concrete example, \figref{fig:full-stencil-example} illustrates the lifetime of the If-Leq (var-const) stencil (the green node in \figref{fig:fibonacci-example-callgraph}) used in the Fibonacci example. The stencil generator is a templated C++ function that generates various stencil variants for the logic shape \code{if (lhs op rhs)} where \code{op} is a comparison operator and \code{lhs} and \code{rhs} has a simple shape. The C++ template instantiation where \code{lhs} is an \code{int} local variable, \code{rhs} is an \code{int} constant, and \code{op} is \code{<=} yields the conceptual C++ logic of the stencil shown in \figref{fig:stencil-example-a}. The logic contains four holes: the offset of the local variable in the stack frame, the constant literal, and two continuations for the true and false branches. Each hole is identified by an ordinal, so at runtime we can specify its desired value. Clang compiles the logic and generates the object code shown in \figref{fig:stencil-example-b}, where the holes are identified by linker relocation records. The MetaVar compiler can figure out the ordinal of each hole in the object file, because the C macro trick we used to create holes associates each hole ordinal to an unique external variable, which translates to a unique symbol referenced by the linker relocation records. The MetaVar compiler parses the object file, and prints out the code shown in \figref{fig:stencil-example-c} to construct the C++ Stencil struct (as defined in \figref{fig:binary-stencil-struct}) for this stencil. The code is compiled and becomes part of the stencil library. At runtime, when we compile the Fibonacci AST, this stencil is selected to implement the if-branch in the Fibonacci function, its holes are filled with concrete values (for example, the local variable \code{n} has offset 8 in the stack frame), and its tail jump instruction is removed and becomes a fallthrough to the continuation (the logic in the true branch), as shown in \figref{fig:stencil-example-d}. This generates the eighteen bytes of executable code in \code{[0x20, 0x32)} of \figref{fig:fibonacci-example-assembly}. Note that, most of the above work happens at library build time. The only work that happens during compilation at runtime is matching the AST with this stencil through a tree pattern matching, a hash table lookup to retrieve the stencil, a {\tt memcpy} of those eighteen bytes, and a few scalar additions to patch the holes, which are all cheap operations.

\begin{figure}
    \centering
    \includegraphics[width=\linewidth]{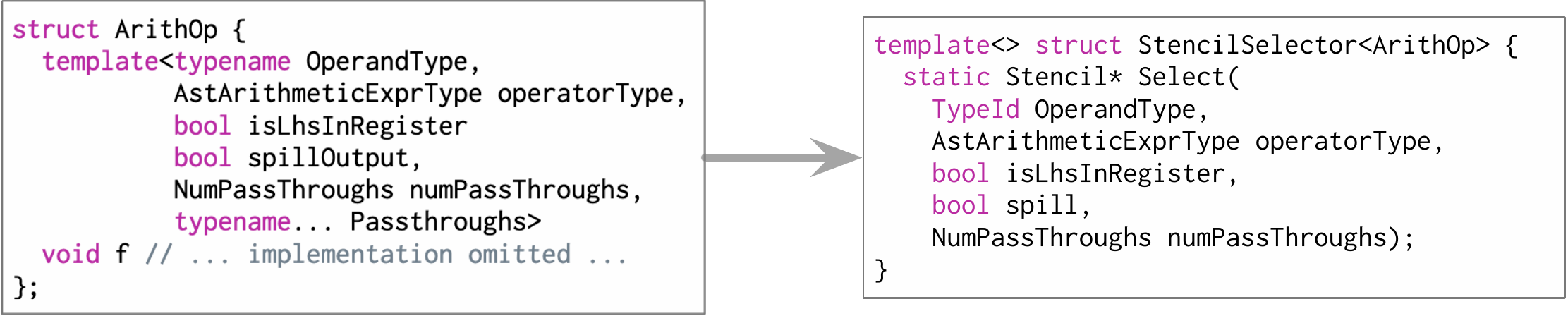}
    \caption{
    An example of the API between the stencil library and the \cp{} runtime. The left side is the stencil generator that generates the various stencil variants of the \code{ArithOp} stencil class. The API on the right side is automatically generated by the MetaVar compiler from the definitions on the left side and exposed to the \cp{} runtime.
        \label{fig:stencil-api}
    }
\end{figure}

Finally, we give an example of the API interface between the stencil library and the \cp{} code generation runtime in \figref{fig:stencil-api}. In the example, the stencil \code{ArithOp}'s C++ template parameters contain the type of the operands, the arithmetic operation kind, etc. MetaVar then creates an API, where the C++ template parameters are converted to function parameters. \begin{wrapfigure}{r}{0.28\linewidth}
    \includegraphics[width=\linewidth]{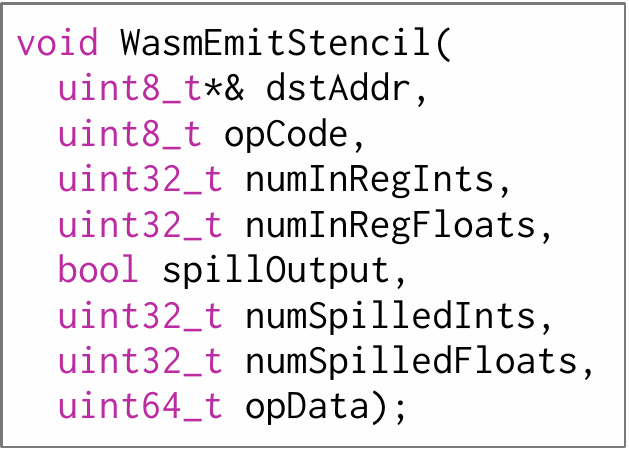}
    \caption{
        The API that handles the codegen for most of the WebAssembly opcode.
        \label{fig:wasm-api}
    }
\end{wrapfigure}The \code{typename} template parameter becomes a special \code{TypeId} enum, while all enum and boolean template parameters are converted unchanged. At runtime, \cp{} selects the stencil by calling the \code{StencilSelector<ArithOp>::Select} function and supplying the runtime-known configuration values. The \code{Select} function uses the parameters as the key to lookup the hash table and returns the stencil. The runtime may then invoke the \cp{} API described in \figref{fig:binary-stencil-struct} to specify the desired values for the holes and generate executable code. The result is a flexible and powerful interface capable of handling the complex language features required for a high-level language.

For WebAssembly, since the bytecode instructions are very low-level and operate on a stack machine, we can further unify the APIs to a single interface for even faster code generation, as shown in \figref{fig:wasm-api}. The \code{opCode}, \code{numInRegInts}, \code{numInRegFloats}, \code{spillOutput} together determines the stencil variant, and the \code{numSpilledInts}, \code{numSpilledFloats} and \code{opData} determines the patch values. This lets \cp{} process most of the WebAssembly instructions (all except control instructions) through this single function API. And all this function does is to look into an array to retrieve the stencil, and then do \cp{} to instantiate this stencil to the destination address. There is not even a switch case on the opcode or on the register configuration. This is why our code generator runs so fast.

\section{Evaluation: WebAssembly Baseline Compiler}
\label{sec:webassembly}

We evaluate our claim that our \cp{}-based WebAssembly compiler achieves significantly lower startup delay and better execution performance compare with the conventional baseline compiler design, and that it narrows the execution performance gap between baseline compilers and optimizing compilers. 

\subsection{Methodology}
\label{sec:webassembly-methodology}

The experiments are run on a single-socket 8-logical-core Intel i7-7700HQ CPU at 2.80GHz with turbo boost on, running Ubuntu 20.04. The machine has 32GB RAM: large enough so that nothing is swapped out. All experiments are repeated at least \num{30} times and the average is reported. Unless otherwise noted, the width of the $95\%$ statistical confidence interval (CI95) is less than $2\%$ of the average absolute performance. This implies that with $95\%$ probability, our reported performance is within $1\%$ relative difference from the truth. Exceptions are \figref{fig:wasm-execution-polybench} and \figref{fig:wasm-execution-polybench-avg}, in which we only report the average performance across three runs, and CI95 is not reported: this is due to the high amount of human labor and computation resource required to run that benchmark.

We evaluate the startup delay and the execution performance of our compiler against eight WebAssembly compilers used in four industrial software products, including three baseline compilers and five optimizing compilers:
\begin{itemize}
    \item Two compilers from Google Chrome 81's V8 Engine: the Liftoff~\cite{liftoffBlog} baseline compiler and the TurboFan~\cite{liftoffBlog} optimizing compiler.
    \item Three compilers from Wasmer 1.0~\cite{wasmer}: the Wasmer SinglePass~\cite{wasmerSinglepassJit} baseline compiler, and two optimizing compilers named Wasmer Cranelift~\cite{wasmerCraneliftJit} and Wasmer LLVM~\cite{wasmerLLVMJit}, using CraneLift~\cite{cranelift} and LLVM~\cite{llvmIrPaper} as backend respectively.
    \item Two compilers from Wasmtime 0.26~\cite{wasmtime}: the Lightbeam~\cite{wasmtimeLightbeam} baseline compiler and an optimizing compiler~\cite{wasmtimeCranelift} using Cranelift.
    \item One compiler from WAVM~\cite{wavm}: an optimizing compiler that uses LLVM~\cite{llvmIrPaper} as backend.
\end{itemize} 

Wasmer LLVM and WAVM both use LLVM as backend, but Wasmer LLVM performs strictly worse than WAVM in both code generation time and execution time on all benchmarks we tested, so we removed Wasmer LLVM from the results. We are also unable to report the numbers for Wasmtime Lightbeam, because it reports an ``unsupported'' error or crashes on all benchmarks we tested. We report code generation time and execution performance for \cp{} and the remaining six compilers. For in-browser compilers (Google Chrome's Liftoff and TurboFan), code generation time measures the time to execute Javascript ``\code{new WebAssembly.Module(data)}'', after the module is fully read into the array \code{data}, to avoid any overhead related to the disk, network, or Javascript. We use browser developer flags to select the desired WebAssembly compiler implementation, following the official instruction~\cite{v8WasmFlagsBlog}. For the other non-browser WebAssembly compilers, we first read the whole WebAssembly module to memory to avoid any disk overhead, then measure the code generation time by timing the C API provided by the respective compiler that compiles a module to executable code. Some of the compilers support multi-threaded code generation. To make the performance results easier to understand, we limit all implementations to use only one CPU in the code generation time benchmarks. However, we note that WebAssembly is designed so that compilation is trivially parallelizable at function level, so our implementation can be easily extended to support multi-threaded compilation with the same scalability as the other compilers, and all conclusions should still hold in a multi-threaded environment. We additionally note that our single-threaded implementation is faster than all benchmark rivals using 8 CPUs.

We benchmarked the compilers on four benchmarks. Two of the benchmarks measure both the execution performance and the startup delay of the compilers: 
\begin{itemize}
    \item CoreMark 1.0~\cite{coremark}, an industrial-standard benchmark to measure the performance of a CPU. We used the default random seed and settings of the benchmark. We confirmed that all compilers produced the correct final checksum.
    \item PolyBenchC 4.2.1~\cite{polybenchc}, the benchmark used in the original WebAssembly paper~\cite{wasmPaper}. It consists of 30 numerical computation kernels extracted from operations in various application domains.
\end{itemize} 
The other two benchmarks test the most important use case of baseline compilers: compiling real-world WebAssembly modules that are very large. We used the following real-world modules:
\begin{itemize}
    \item AutoCAD Web App~\cite{autocadWebApp}, a \SI{47.5}{\mega\byte} WebAssembly module containing a \SI{40.2}{\mega\byte} WebAssembly code section. The module is downloaded directly from the app website using its public URL. Notably, Liftoff also used this as a benchmark~\cite{liftoffBlog}.
    \item \textit{clang.wasm}~\cite{clangDotWasm}, a \SI{30.5}{\mega\byte} WebAssembly module containing a \SI{27.5}{\mega\byte} WebAssembly code section. The module is downloaded from the official mirror.
\end{itemize} 
For these two benchmarks, we only measure the code generation time, as there is not a clear criterion to quantitatively measure their execution performance.

\subsection{Code Generation Performance}

\begin{figure}
    \begin{minipage}{0.53\textwidth}
        \vspace{-1em}
        \includegraphics[width=\linewidth]{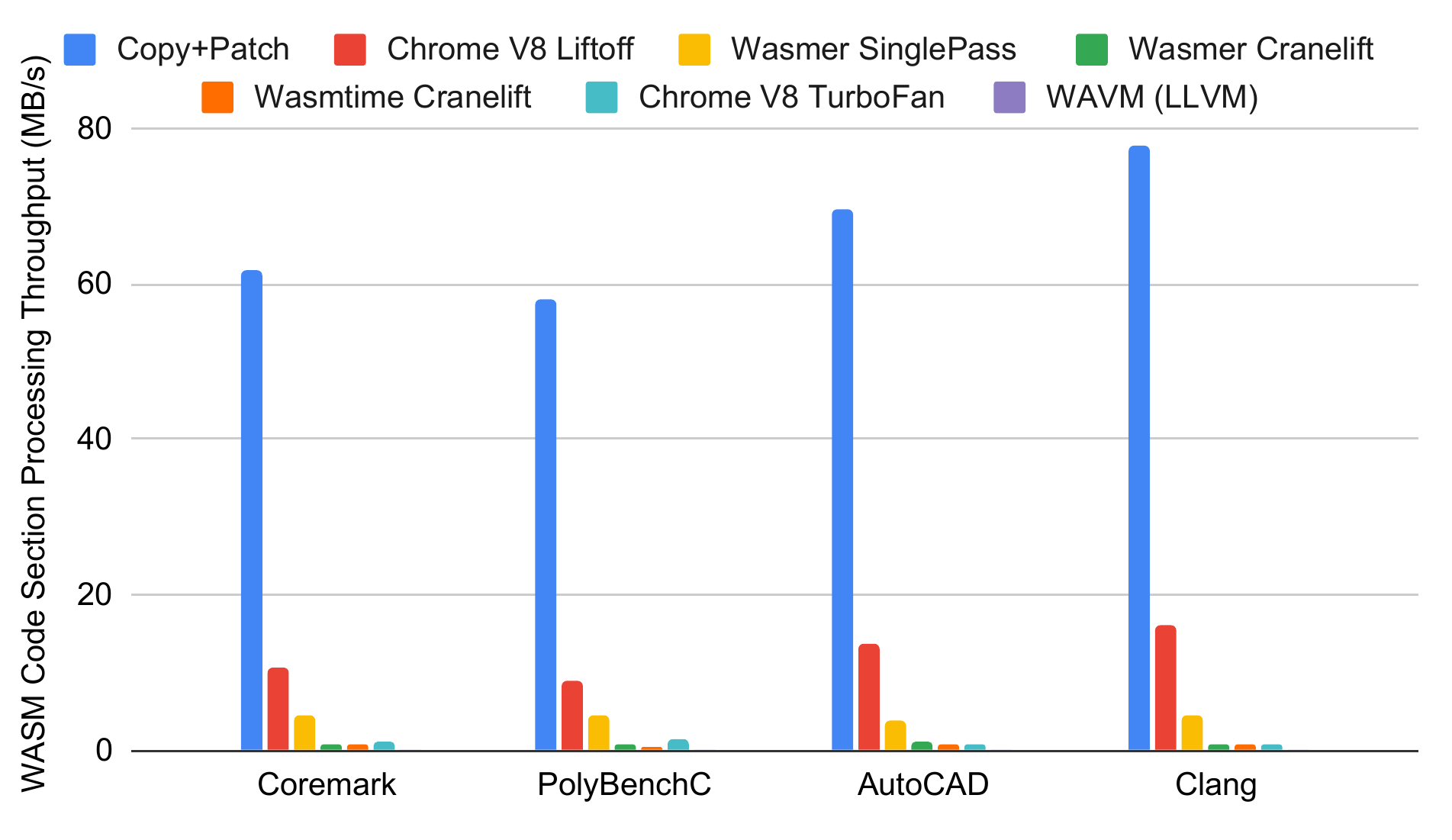}
        \caption{
            The absolute throughput of each WebAssembly compiler, in terms of megabytes of WebAssembly code section processed per second. The PolyBenchC column records the average throughput of the 30 PolyBenchC benchmark modules (higher is better). The 95\% confidence intervals are the same as in \figref{fig:wasm-codegen-normalized}.
            \label{fig:wasm-codegen-absolute}
        }
    \end{minipage}
    \hfill
    \begin{minipage}{0.45\textwidth}
        \vspace{-1em}
        \includegraphics[width=\linewidth]{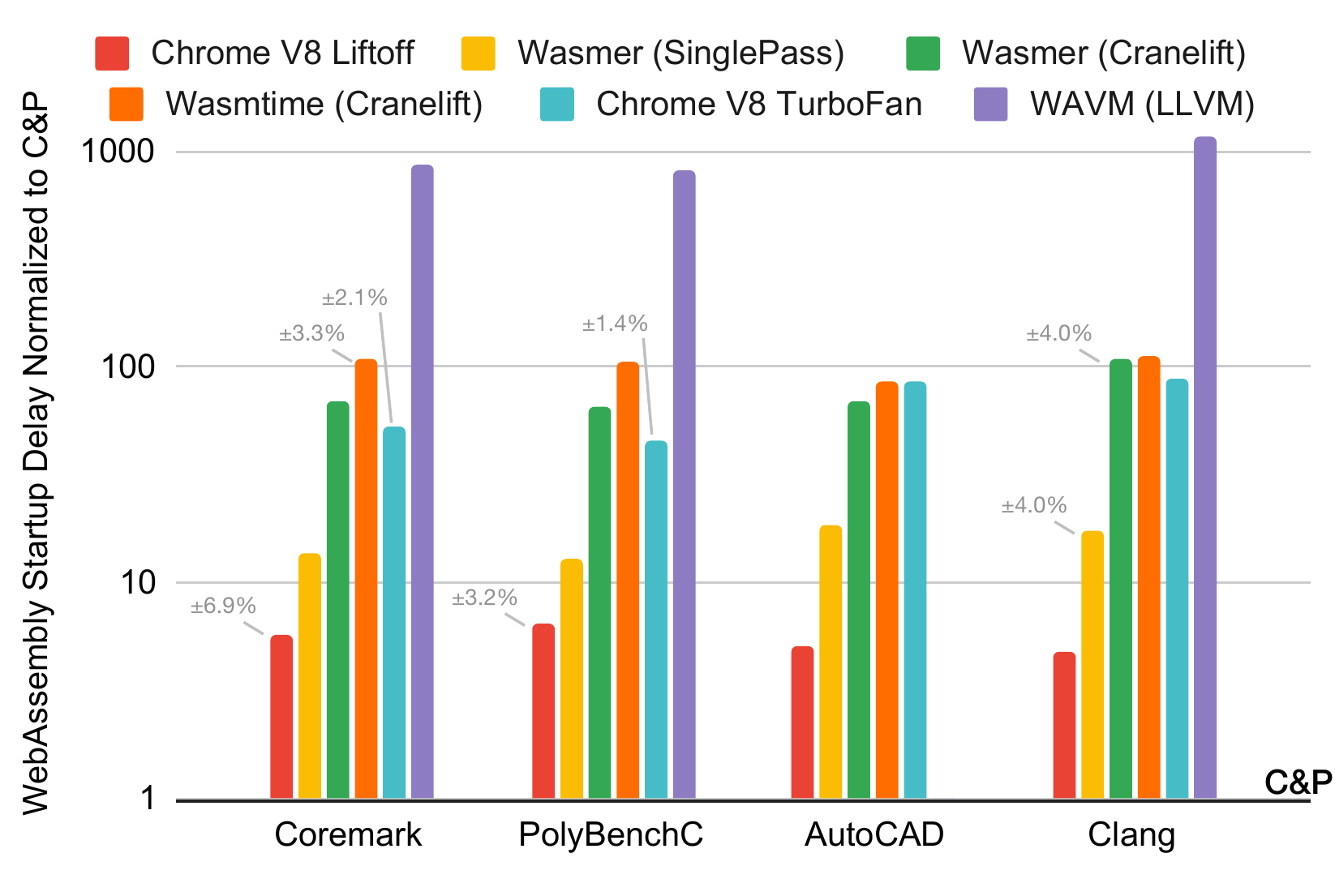}
        \caption{
            The log-scale normalized startup delay of each WebAssembly compiler, with \cp{} normalized to 1. The PolyBenchC column records the average  startup delay of the 30 PolyBenchC benchmark modules. All 95\% confidence intervals greater than $\pm 1\%$ are shown.
            \label{fig:wasm-codegen-normalized}
        }
    \end{minipage}
    \vspace{-1em}
\end{figure}

We measured the startup delay of all compilers on all benchmarks. The AutoCAD WebAssembly module appears to trigger a bug in LLVM, causing all LLVM-based compilers (WAVM and Wasmer LLVM) to enter a dead loop (did not finish in 4 hours). All other compilations are successful. \figref{fig:wasm-codegen-absolute} shows the absolute throughput of each compiler, in terms of megabytes of WebAssembly code processed per second. 
Since the high throughput of \cp{} renders the throughput bar of many compilers barely visible, we plot the normalized log-scale startup delay (with \cp{} normalized to 1) in \figref{fig:wasm-codegen-normalized}.

As shown in \figref{fig:wasm-codegen-normalized}, the startup delay of our compiler is consistently lower than all the other compilers. Compared with baseline compilers, it is \num{4.9}\X{}--\num{6.5}\X{} faster than Liftoff, and \num{12.7}\X{}--\num{18.5}\X{} faster than Wasmer SinglePass. Compared with optimizing compilers, the startup delay is two to three orders of magnitudes lower. 

\begin{wrapfigure}[13]{r}{0.24\linewidth}
    \vspace{-2.5em}
    \includegraphics[width=\linewidth]{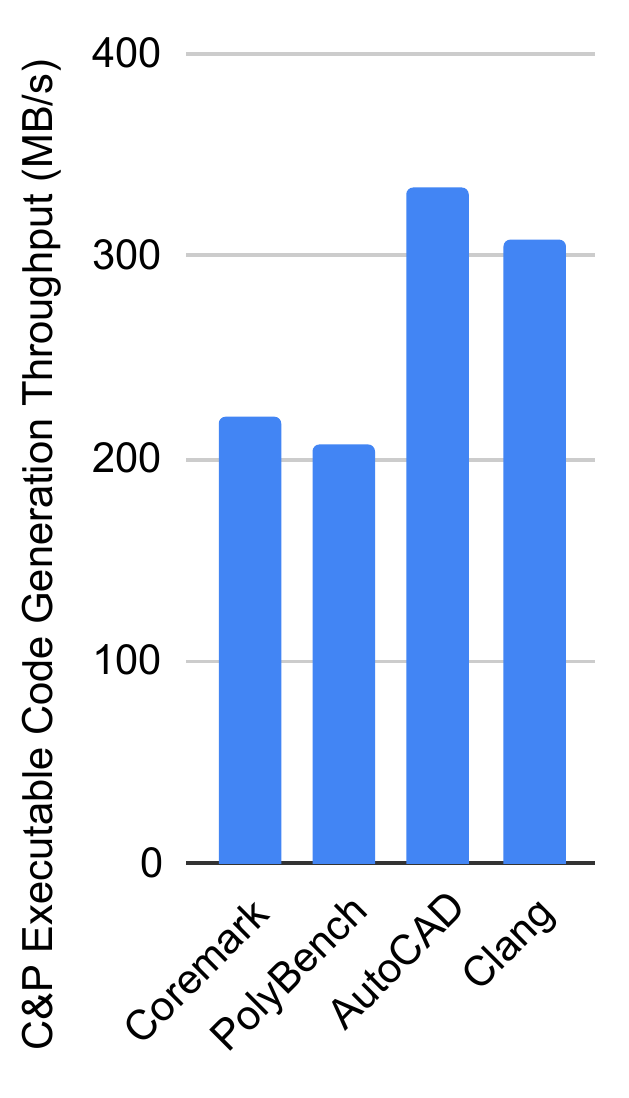}
    \vspace{-3em}
    \caption{
        Throughput of machine code generated by \cp{} (MB/s).
        \label{fig:wasm-codegen-binary-code-throughput}
    }
\end{wrapfigure}

The code generation throughput of our compiler on Coremark and PolybenchC is lower than on AutoCAD and Clang. This is because the code generation memory manager in our implementation takes about \SI{0.2}{\milli\second} to initialize. This cost is negligible for large modules, but the Coremark and PolyBenchC benchmark modules are small, containing only about \SI{30}{\kilo\byte} of WebAssembly code. \cpcap{} needs \SI{0.6}{\milli\second} to generate code for one such module, so the \SI{0.2}{\milli\second} initialization cost takes one third of the time.

\figref{fig:wasm-codegen-binary-code-throughput} demonstrates the throughput of machine code generated by \cp{}. As one can see, on large modules like AutoCAD and Clang, we are capable of generating more than 300MB of machine code per second using a single CPU. Such throughput is made possible by \cp{}'s design that turns the task of code generation into, literally, copy (by memcpy) and patch (by a few scalar additions).

\subsection{Execution Performance}
\label{sec:webassembly-exec}

\begin{figure}
    \begin{minipage}[b]{0.209\textwidth}
        \includegraphics[width=\linewidth]{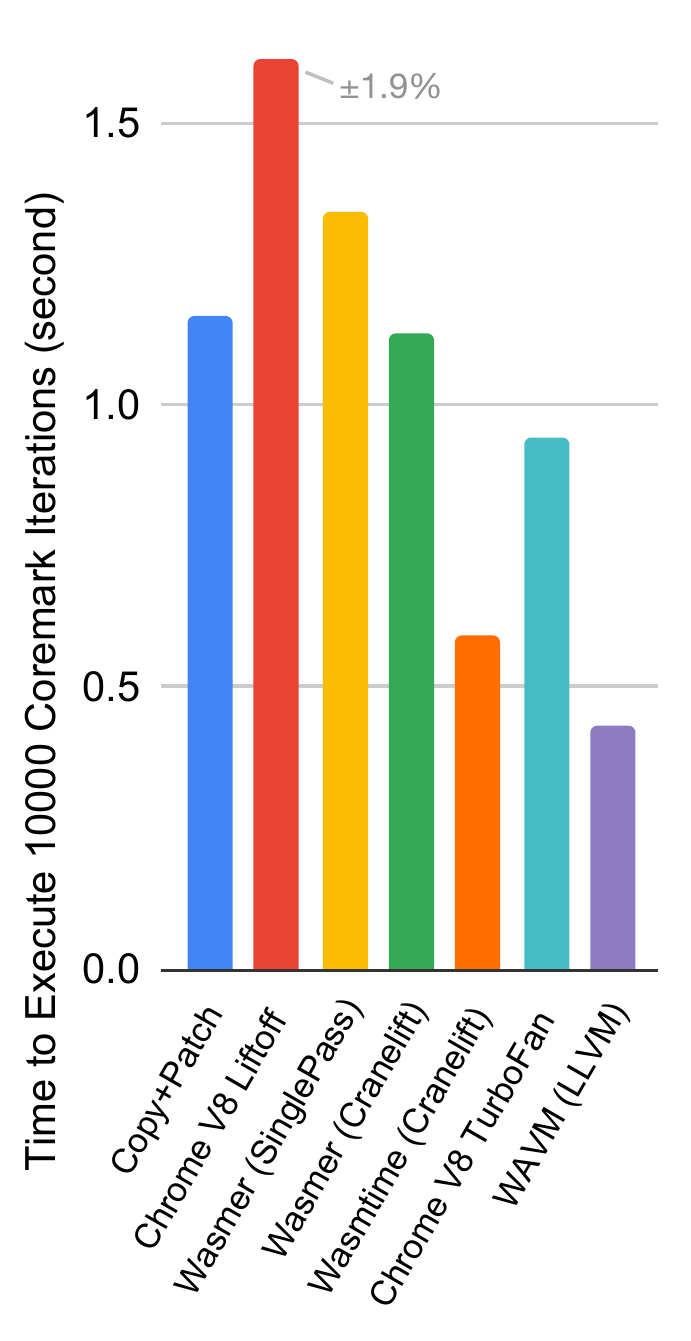}
        \caption{
            WebAssembly exec time, Coremark.
            \label{fig:wasm-execution-coremark}
            \vspace{-1em}
        }
    \end{minipage}
    \hfill
    \begin{minipage}[b]{0.6\textwidth}
        \includegraphics[width=\linewidth]{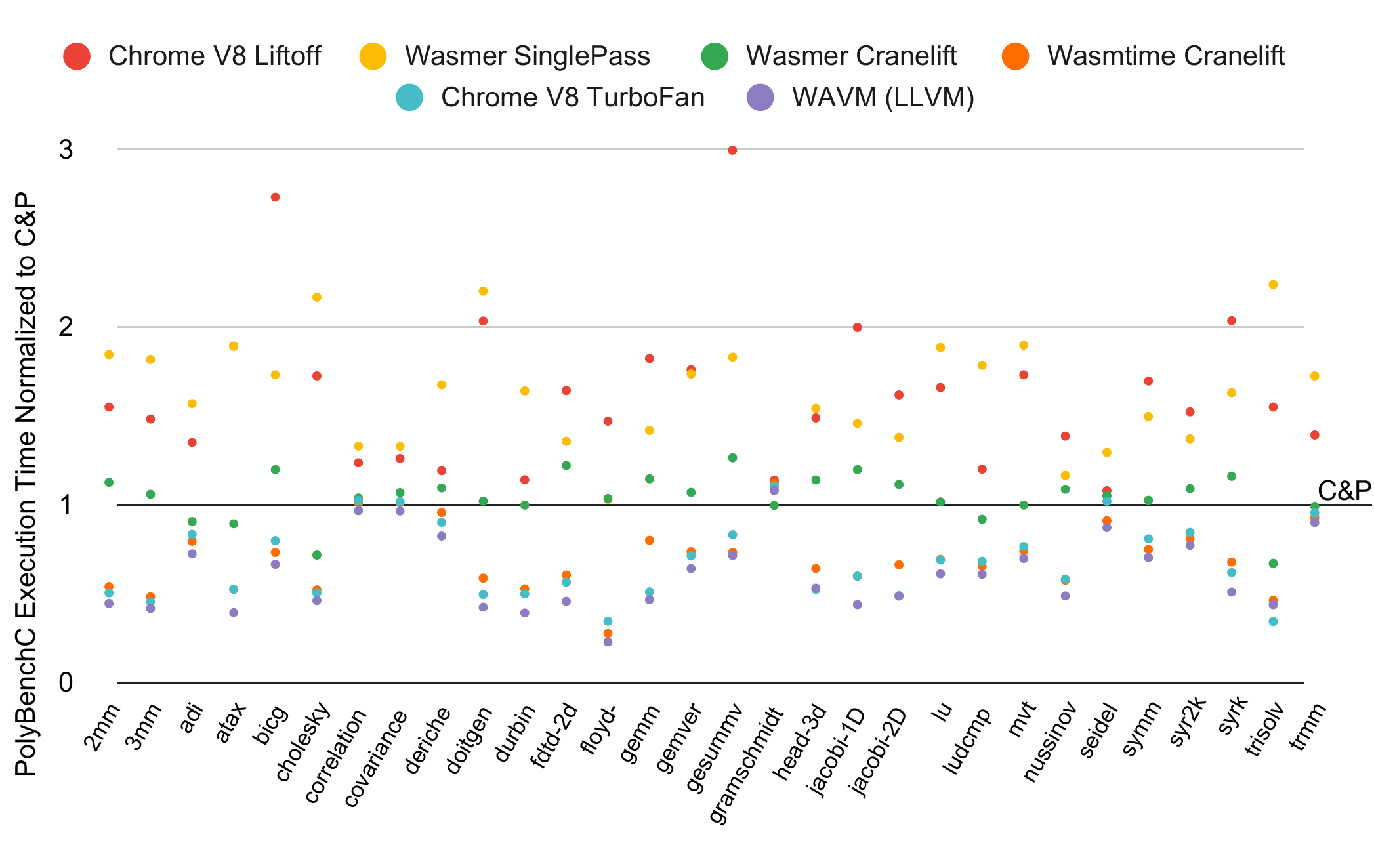}
        \caption{
            WebAssembly normalized execution time on each of the 30 PolyBenchC benchmarks, normalized to \cp{}.
            \label{fig:wasm-execution-polybench}
            \vspace{-1em}
        }
    \end{minipage}
    \hfill
    \begin{minipage}[b]{0.176\textwidth}
        \includegraphics[width=\linewidth]{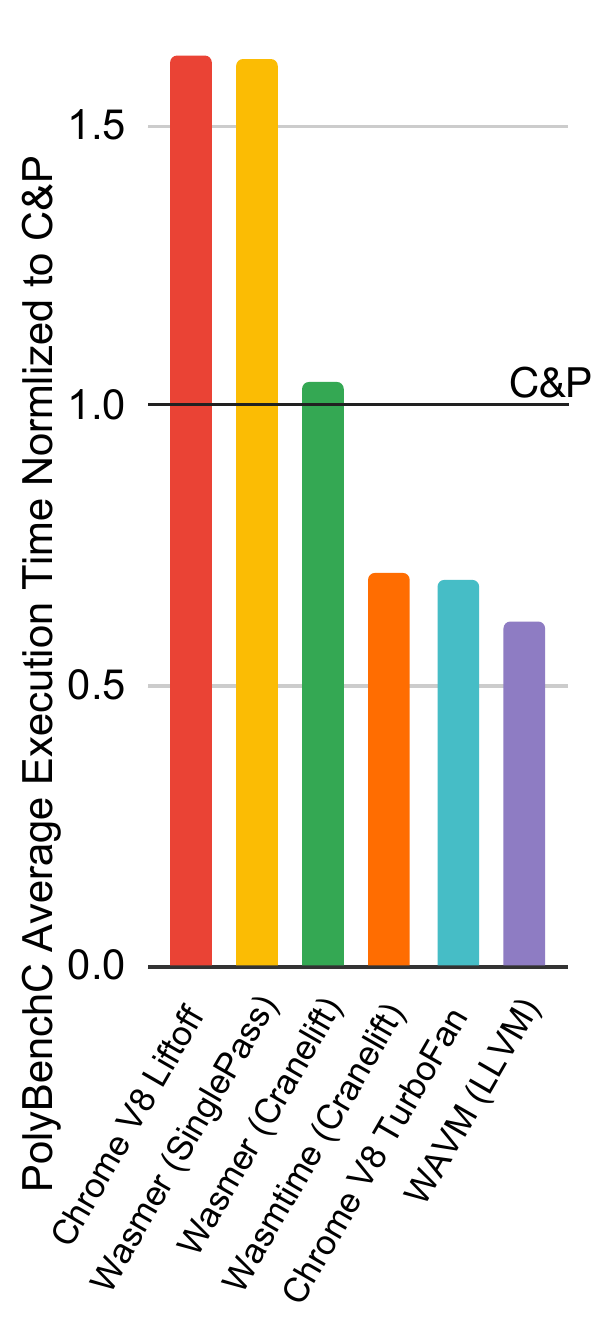}
        \caption{
            Average of \figref{fig:wasm-execution-polybench}.
            \label{fig:wasm-execution-polybench-avg}
            \vspace{-1em}
        }
    \end{minipage}
\end{figure}

\cpcap{} not only generates code fast, but also generates fast code. We measure the execution performance of all compilers on the Coremark and PolyBenchC benchmarks.
\figref{fig:wasm-execution-coremark} reports the time to execute 10000 Coremark iterations. \figref{fig:wasm-execution-polybench} reports the normalized execution time of each of the 30 PolyBenchC benchmarks, with \cp{}'s execution time normalized to 1. The average normalized execution time of PolyBenchC is reported in \figref{fig:wasm-execution-polybench-avg}.

As one can see, on Coremark and all 30 PolyBenchC benchmarks, \cp{} consistently performs better than all baseline compilers. Compared with Liftoff, we are on average $39\%$ faster on Coremark, and on average $63\%$ faster on PolyBenchC. 
The speedup compared with Wasmer SinglePass is similar. The performance of \cp{} is even comparable with one optimizing compiler (Wasmer Cranelift), being $2.6\%$ slower on Coremark but $4.6\%$ faster on PolyBenchC. 
Therefore, we conclude that \cp{} replaces the role of baseline compilers, and narrows the performance gap between baseline compilers and optimizing compilers.

\subsection{Other Metrics}

Our WebAssembly implementation contains \wasmStencilNum{} stencils taking \wasmStencilSize{} of memory, which is built in less than a minute at compiler installation time. The stencil library is small because we did not implement supernodes (e.g., arithmetic operations taking a constant operand, branch if a relational operation is true). We anticipate that our execution performance could be further improved, at a cost of a larger stencil library and a slightly higher startup delay, if we had implemented supernodes.

\section{Evaluation: High-Level Language Compiler}
\label{sec:evaluation}

In this section, we evaluate our C-like high-level language compiler built on top of copy-and-patch. The compiler is implemented as a domain-specific language (DSL) library embedded in the C++ host language. The language constructs are exposed to users as a set of C++ APIs (for example, \code{Declare}, \code{For}, \code{If}, \code{Call}), which build the user program AST behind the scene. \figref{fig:fibonacci-example-ast} demonstrates an example program in our DSL. Once a user program is built, it may be interpreted using the AST interpreter, compiled to binary code using \cp{}, or lowered to LLVM IR and then compiled to binary code using LLVM. The three backends are functionally equivalent and operate on the same input program. This allows an apple-to-apple comparison of the startup delay and execution performance tradeoffs presented by the techniques.

Our compiler is a general purpose compiler since it compiles a C-like language. \secref{sec:pareto-frontier} evaluates its performance on a few well known microbenchmark algorithms. However, one of the most important use cases of such a compiler is in SQL databases, where SQL queries are lowered to concrete program logic, then compiled to binary code using the compiler and executed. Therefore, we implemented a SQL query compiler prototype on top of our compiler and evaluated its performance on TPC-H queries, which is  elaborated in \secref{sec:database-benchmarks}.

In both cases, we evaluate our central claim that \cp{} replaces both \llvm{} \ozero{} compilation and interpreters. \cpcap{} generates code that executes faster than \ozero{} compilation and an order of magnitude faster than interpretation, with negligible startup delay. Higher \llvm{} optimization levels are still desirable for long-running computations, but \cp{} narrows the range in which they are useful, as their compilation overhead is three orders of magnitude higher than \cp{} code generation. We also quantify the effect of the \cp{} optimizations, to shed light on where the performance of its generated code comes from.



\subsection{Methodology}

The hardware environment is the same as described in \secref{sec:webassembly-methodology}. All experiments are repeated at least \num{100} times and the average is reported. For all experiments, the width of the $95\%$ statistical confidence interval (CI95) is less than $2\%$ of the average absolute performance. This implies that with $95\%$ probability, our reported performance is within $1\%$ relative difference from the truth. Time is measured using the {\tt clock\_gettime()} Linux high resolution clock API. The stencils and the \cp{} runtime are compiled by Clang++ 10 using options {\tt -O3} {\tt -DNDEBUG}. The MetaVar system and the LLVM backend of our metaprogramming language use LLVM library v10.0.0, the latest version at the time of writing. LLVM has an additional fixed startup cost of about \SI{1}{\milli\second} when compiling a module. Therefore, for microbenchmarks where the modules are small, we amortized out that cost by generating \num{100} clones of the function inside the module and report $1/100$ of total code generation time, to reflect the true time \llvm{} spent generating code. 

We implemented three microbenchmarks and eight relational queries from the standard TPC-H database management system benchmark suite~\cite{tpch}. We generate the TPC-H benchmark database using the official generator {\tt dbgen} with a scale factor of \num{0.3} (about 380MB data). This scale factor is a typical size of the database partition assigned to each CPU when TPC-H is used to benchmark a distributed database~\cite{memsqlPersonalCommunication,postgresTpch}. To emphasize how LLVM optimization levels and interpreters compare to \cp{}, we report their startup and execution times normalized to multiples of \cp{}.
We report the absolute running time for every experiment in \appendixNumbers{} in the supplemental material.

\subsection{Microbenchmarks}
\label{sec:pareto-frontier}

We explore the startup--execution time Pareto frontier of several approaches to online code generation. The Pareto frontiers are the set of Pareto efficient points for which no gain can be had in startup delay without giving up some execution time and vice versa. We compare \cp{} (\cpshort{}) to the \llvm{} compilation levels \ozero{}, \oone{}, \otwo{}, and \othree{}. We also include the Peloton interpreter~\cite{pelotonInterpreter} from the query execution engine of the Peloton database management system~\cite{peloton, pelotonQueryCompilation}. It interprets a subset of the LLVM IR, but this IR is low-level and leads to high interpretation cost. Therefore, to get a better baseline for interpretation, we developed a higher-level AST Interpreter that runs approximately \num{1.5}\X{} faster than Peloton's interpreter.

We used these systems to execute three microbenchmarks and recorded their startup delay and execution time on synthetic input. The microbenchmarks include a function to compute Fibonacci numbers (\figref{fig:fibonacci-example-ast}), an implementation of Euler's sieve~\cite{euler_sieve}, and an implementation of quicksort. The Fibonacci function is small with two recursive calls, demonstrating the efficiency of the generated code across calls. Euler's sieve is heavy on arithmetic and demonstrates the efficiency of generated code across compute stencils. And quicksort is a mix of these traits.

The \cpshort{} technique moves the Pareto frontier of the three microbenchmarks, effectively rendering both \ozero{} compilation and interpretation obsolete. \figref{fig:eval-micro} plots the startup delay of each approach against the execution times for the three microbenchmarks. LLVM \otwo{} and \othree{} have the same startup and execution times, so we show only LLVM \otwo{}. All times are normalized to \cpshort{}. In all cases, LLVM \ozero{} falls behind the Pareto frontier, meaning \cpshort{} is a strictly better choice.

The AST interpreter generally has a lower startup cost than \cpshort{}, but in both cases the cost is negligible (\SI{1}{\micro\second} vs \SI{1}--\SI{3}{\micro\second}) compared to all but trivial program executions. We therefore posit that using the \cp{} technique is preferable to interpreters in metaprogramming systems. Moreover, the Pareto frontier line between \cpshort{} and LLVM \code{-O1} is steep, meaning the improved execution performance comes at a high compilation cost. For example, for Euler's sieve, a \num{1442}\X{} higher compilation cost yields only a 24\% execution performance gain. This decreases the number of applications that benefit from optimizing compilation, and it should therefore be reserved for functions that will be run for a significant period of time.

\newcommand{\graphScale}{0.37}
\begin{figure}
    \begin{minipage}[t]{0.05\linewidth}
        \includegraphics[scale=\graphScale]{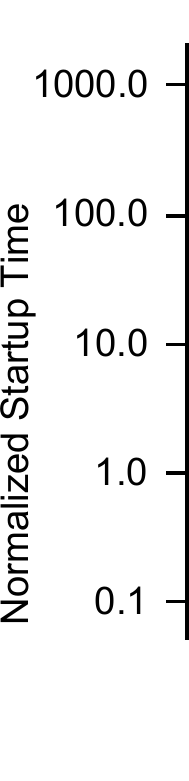}
    \end{minipage}
    \begin{minipage}[t]{0.31\linewidth}
        \includegraphics[scale=\graphScale]{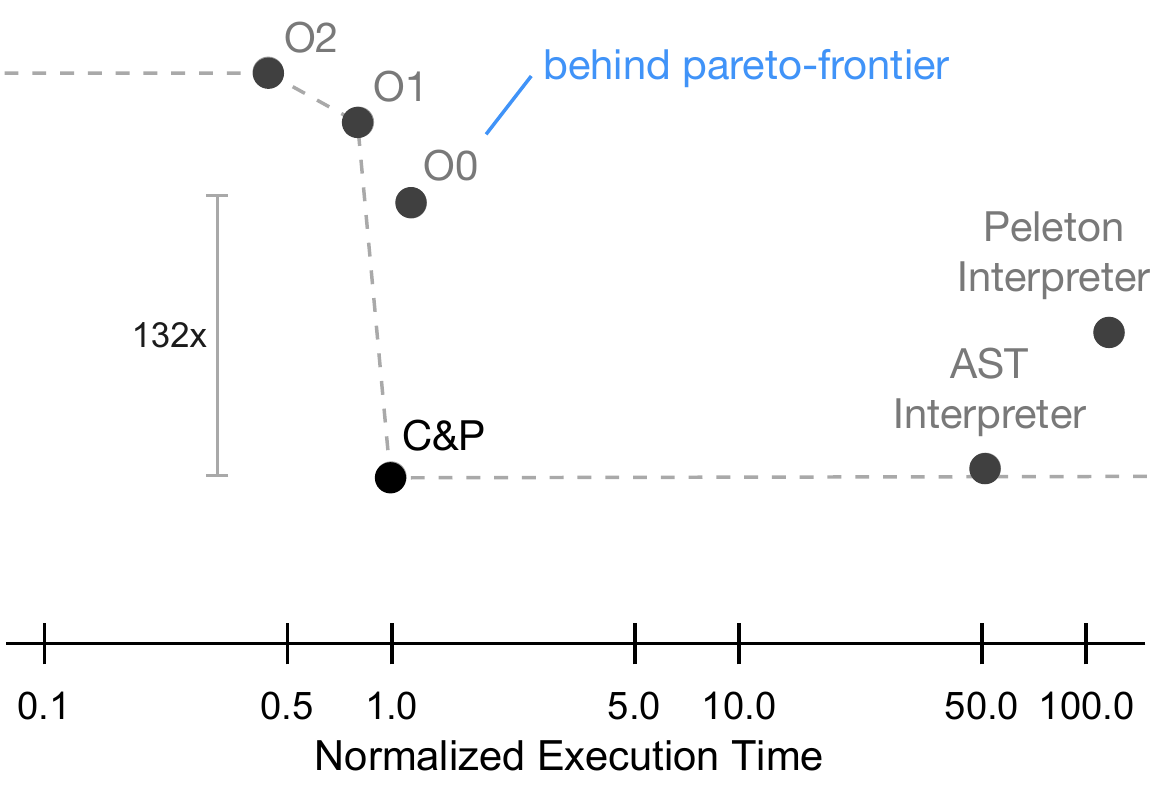}
        \center
        \subcaption{
            Fibonacci numbers
            \label{fig:micro-fib}
            \vspace{-0.5em}
        }
    \end{minipage}
    \begin{minipage}[t]{0.31\linewidth}
        \includegraphics[scale=\graphScale]{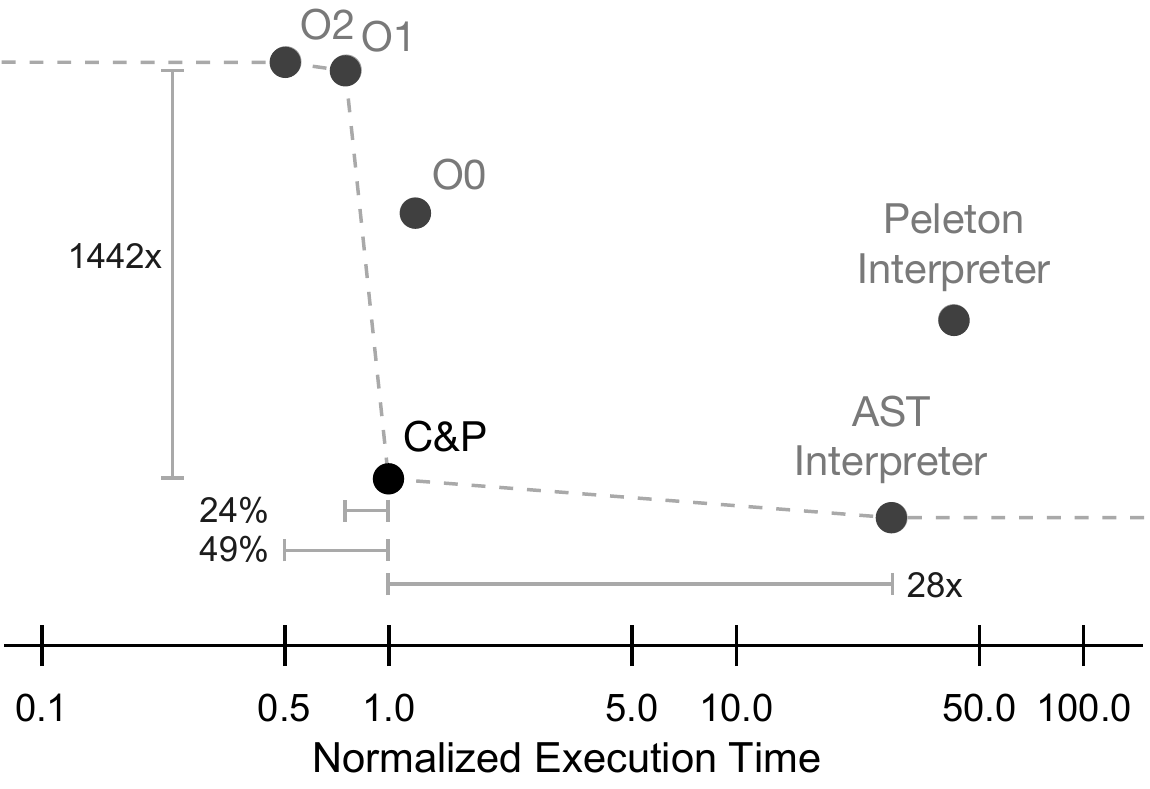}
        \center
        \subcaption{
            Euler's sieve
            \label{fig:micro-sieve}
        }
    \end{minipage}
    \begin{minipage}[t]{0.31\linewidth}
        \includegraphics[scale=\graphScale]{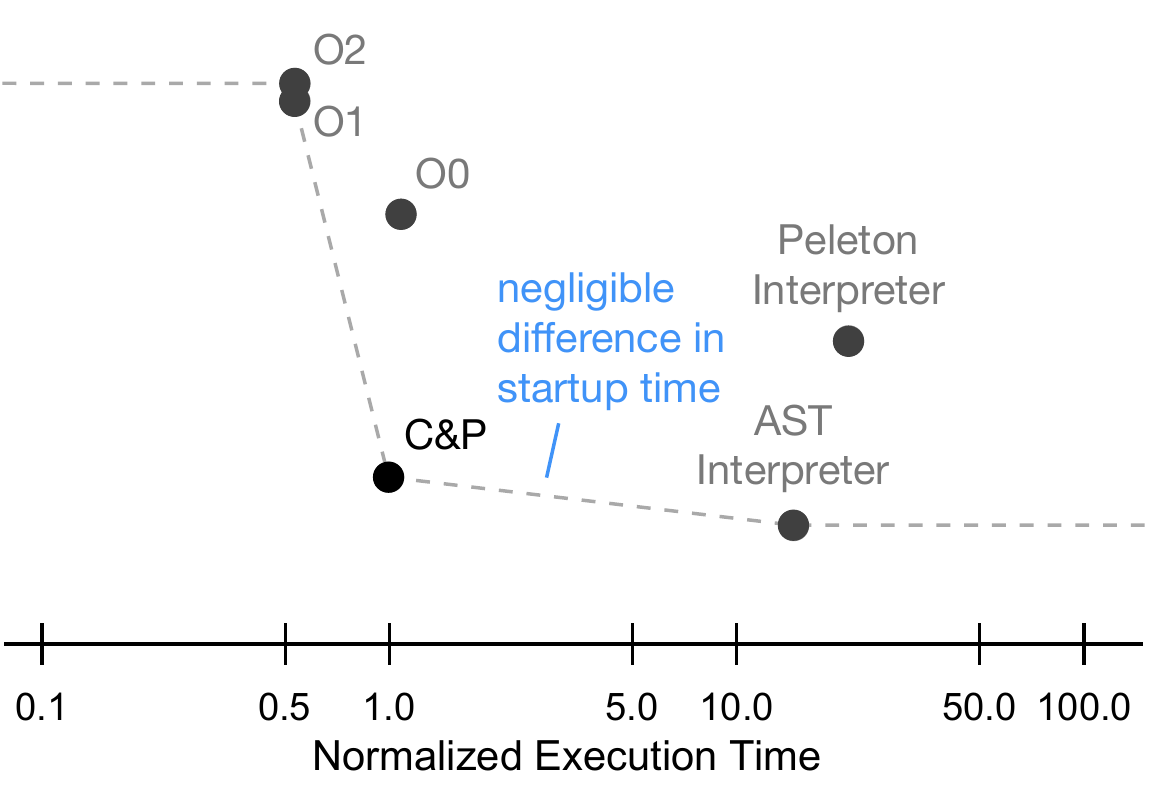}
        \center
        \subcaption{
            Quicksort
            \label{fig:eval-micro-qsort}
        }
    \end{minipage}
    \caption{
        The Pareto frontier of LLVM's compilation levels, \cpshort{}, and interpreters on three microbenchmarks.
        \cpshort{} dominates the \llvm{} \ozero{} optimization level: it produces better code in two orders of magnitude less time. \cpshort{} also replaces interpretation in practice: both have negligible startup overhead, but \cpshort{}'s generated code runs an order of magnitude faster.
        \label{fig:eval-micro}
        \vspace{-1em}
    }
\end{figure}

Finally, we measure the startup and execution time of the microbenchmarks implemented in Java, using OpenJDK 11's Java bytecode interpreter~\cite{java_bytecode_interpreter} and HotSpot JIT~\cite{java_hotspot_jit}. This compares \cpshort{} to an industry-strength interpreter and JIT. We note that this is not an apple-to-apple comparison, since Java has the advantage of working on pre-compiled and pre-optimized bytecode, instead of a high-level AST generated at runtime. Nevertheless, the Java interpreter's execution time is \num{4.4}\X{}--\num{36}\X{} slower than \cpshort{}. The HotSpot JIT's compilation time, measured with JITWatch~\cite{jitwatch_tool}, is from $40\%$ faster to $20\%$ slower than \llvm{} \othree{}, while its execution time is $20\%$--$70\%$ slower than \llvm{} \othree{}. Thus, the conclusion of our comparison between \cpshort{} and \llvm{} also holds for Hotspot JIT.

\subsection{TPC-H Performance}
\label{sec:database-benchmarks}

A typical SQL query compiler, such as Hyper~\cite{neumann2011}, PostgreSQL~\cite{postgresJit},  Peloton~\cite{pelotonQueryCompilation}, or MemSQL~\cite{memsqlCodegen}, consists of three components: the SQL parser, the SQL query planner, and the plan executor. The SQL parser parses a user-provided SQL text to a SQL AST. The SQL query planner determines the most efficient plan to execute the SQL AST, and generates a query plan tree that describes the plan (e.g., join order, join method, operator order, and so on). Finally, the plan executor lowers the query plan tree to LLVM IR, then LLVM is invoked to compile the IR to binary code. 

We have implemented such a database query compiler, but our query compiler lowers the input query plan tree to a program in our C-like language using our metaprogramming DSL library. The program may then be executed using the AST interpreter, the \cp{} backend, or the LLVM backend, letting us compare their performance on database workloads. Our query compiler supports most of the important SQL execution plan nodes, including table scan, filter, hash join, projection, aggregation, group-by, order-by, and a number of SQL scalar operators. However, we note two differences between our prototype and an industrial implementation. 

First, since the SQL parser and the SQL query planner are independent from the code generation techniques, for the purpose of our benchmark, we did not implement these components. The input to our compiler is thus a hand-coded query plan tree, instead of a SQL query text. While the query plan nodes we supported should be sufficient to execute most of the TPC-H queries, many queries require complicated rewriting to yield a reasonable query plan tree, which requires a lot of labor that is unrelated to the main point of our evaluation. Therefore, we only implemented those 8 TPC-H queries whose execution plans follow directly from the query text. 

Second, our implementation is simple, with only 600 lines of core logic (not counting comments, empty lines and curly braces) to call our DSL APIs and construct the program. Industry database management systems are much more complex, and generate much larger programs than we do to support the complexities required by a real-world database. These complexities include SQL-specific semantics (e.g. null-related behavior, collations), transaction semantics, more complex data structure and execution strategy, larger-than-RAM datasets, parallel and distributed execution, and more. Therefore, conditioning on the same compilation technique, an industrial database needs much \textit{more} time to compile a query than our prototype does, which further motivates fast compilation techniques. We validated this using the MemSQL product trial: the TPC-H queries take \num{4.8}\X{}--\num{52}\X{} (average \num{16.4}\X{}) longer to compile than our query compiler when both are using \llvm{} \othree{}, and one compilation can take up to 4.5 seconds, making compilation time a concern. 


\begin{figure*}
    \begin{minipage}[t]{0.49\linewidth}
        \centering
        \includegraphics[width=\linewidth]{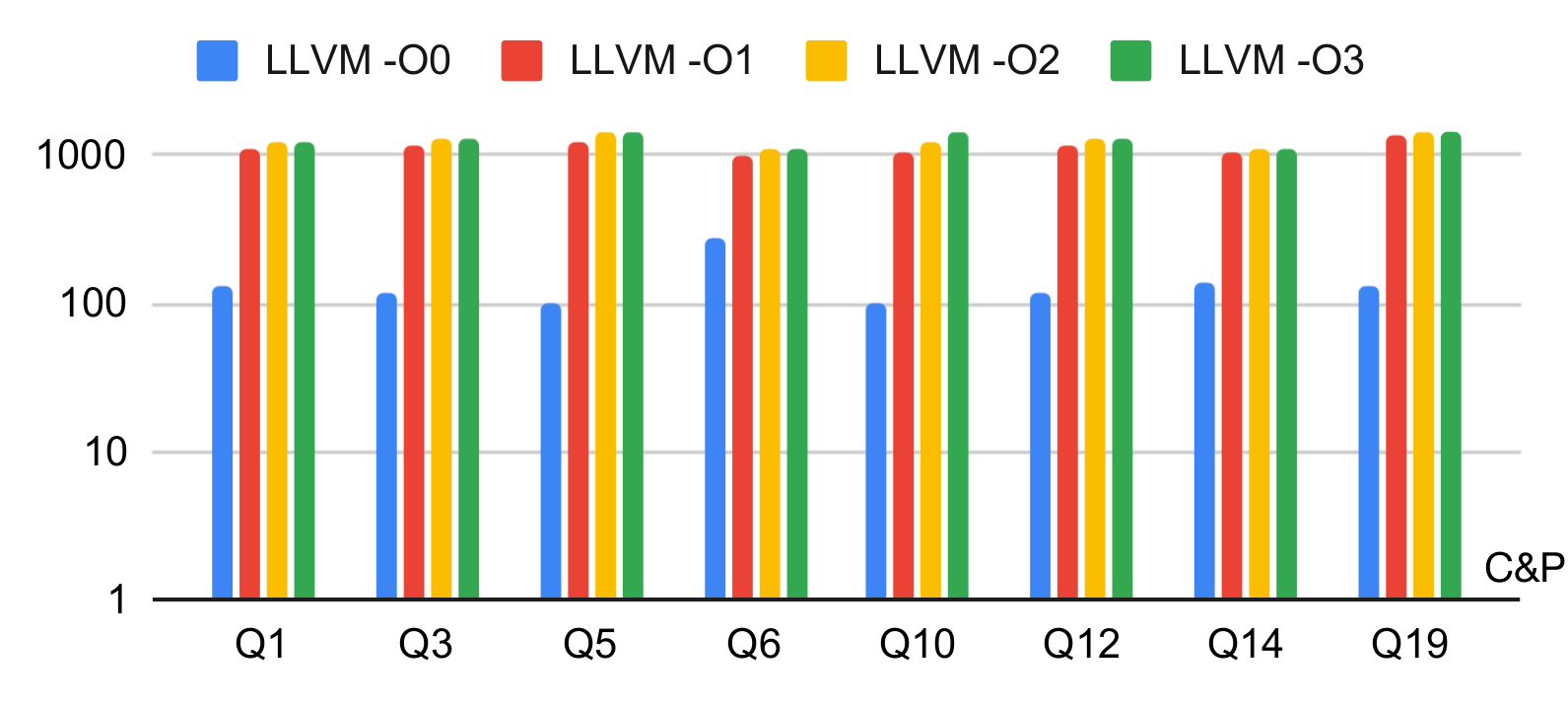}
    \end{minipage}
    \hfill
    \begin{minipage}[t]{0.49\linewidth}
        \centering
        \includegraphics[width=\linewidth]{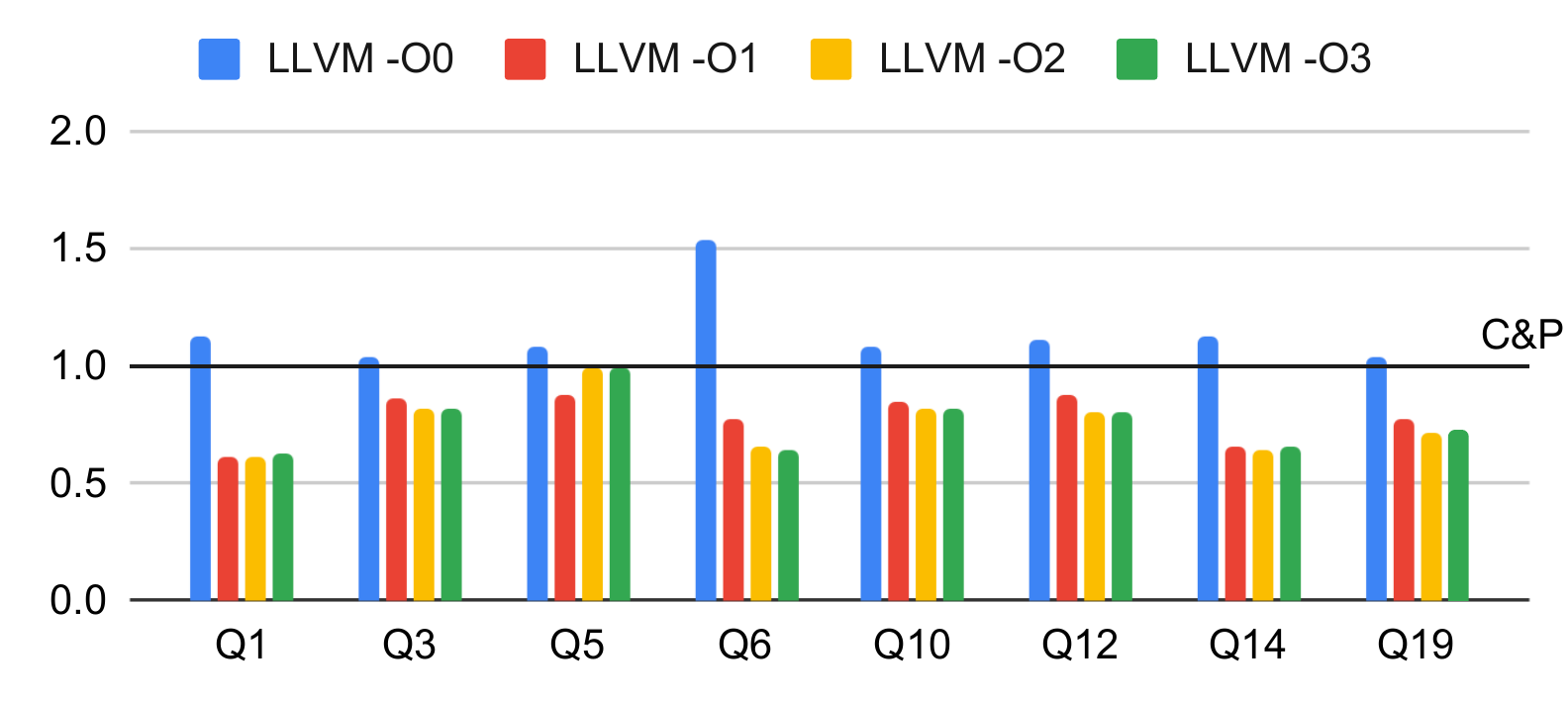}
    \end{minipage}
    \vspace{-1em}
    \caption{
        Normalized startup delay (left) and execution time (right), as multiples of \cpshort{}, of code generated by the LLVM optimization levels across the TPC-H queries. \cpshort{} generates code two orders of magnitude faster (up to 276\X{}) than LLVM \ozero{} and three orders of magnitude (up to 1435\X{}) faster than the other LLVM optimization levels. The resulting code performs on average 14\% better than LLVM \ozero{}, 22\% slower than LLVM \oone{}, 25\% slower than LLVM \otwo{}, and 24\% slower than LLVM \othree{}.
        \label{fig:tpch-llvm}
        \vspace{-1em}
    }
\end{figure*}

\begin{figure}
    \begin{minipage}[t]{0.48\linewidth}
        \includegraphics[width=0.9\linewidth]{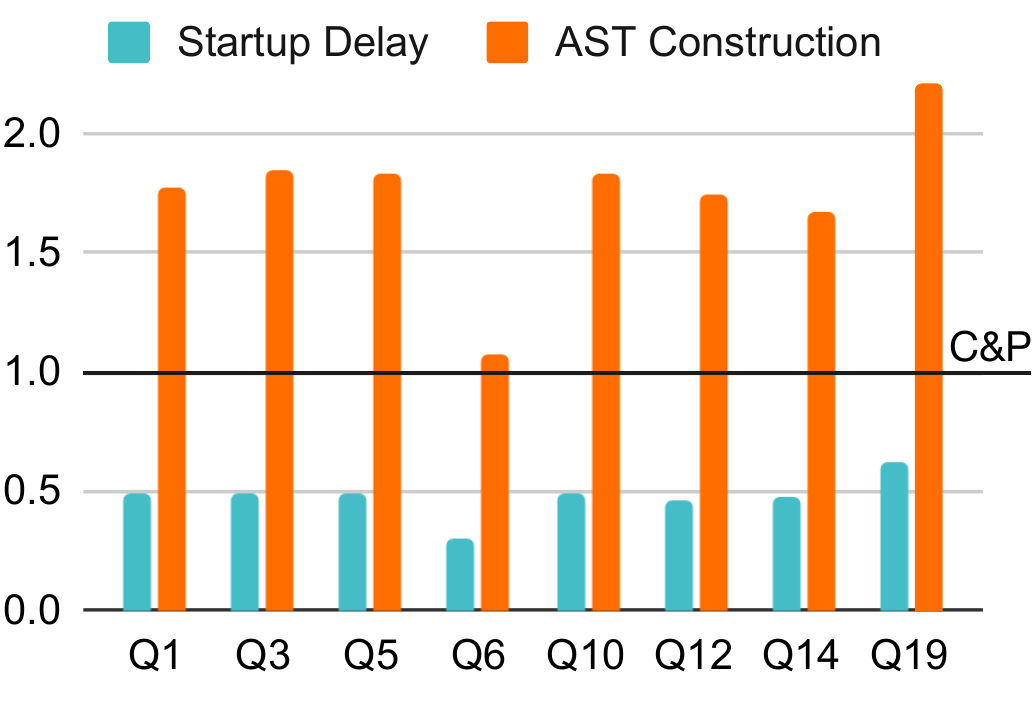}
    \end{minipage}
    \hfill
    \begin{minipage}[t]{0.48\linewidth}
        \includegraphics[width=0.9\linewidth]{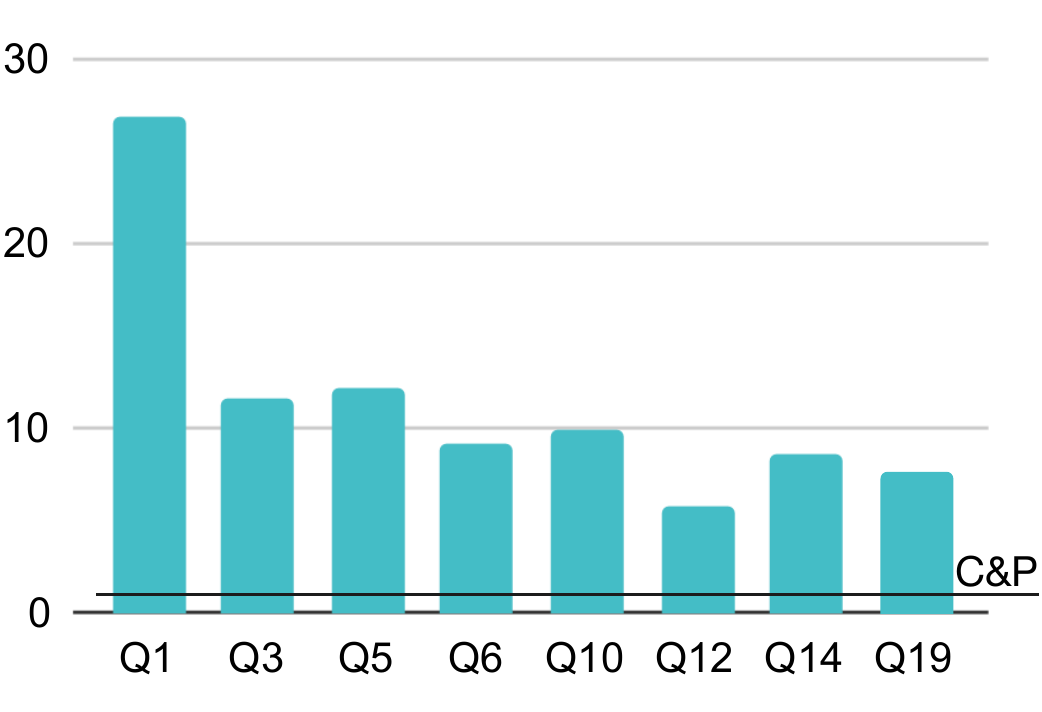}
    \end{minipage}
    \vspace{-1em}
    \caption{
        Normalized startup delay and AST construction cost (left) and execution time (right) of the AST interpreter across the TPC-H queries, as multiples of \cpshort{}. The interpreter has less than half the startup delay of \cpshort{}, but both costs are negligible as it takes longer to construct the AST. And the interpreter executes 6--27 times slower than \cpshort{}.
        \label{fig:tpch-interp}
        \vspace{-1em}
    }
\end{figure}

\subsubsection*{Comparison to LLVM Compilation}

We demonstrate the performance of \cp{} compared to \llvm{} on TPC-H benchmark queries. We build ASTs from query plans using our metaprogramming system and lower them to binary code with \cp{} and \llvm{} at each optimization level. \figref{fig:tpch-llvm} (left) shows the compilation time of each \llvm{} optimization level and \figref{fig:tpch-llvm} (right) shows the execution time of the resulting code, both normalized to multiples of \cp{}. \llvm{} \ozero{} takes two order of magnitude more time to compile and produces less performant code. The other \llvm{} optimization levels take about three orders of magnitude more time to compile (\num{1083}--\num{1435}\X{}), but produces 1.2\%--37.3\% better performing code (the average and median for \othree{} are both 24\%). This shows that the results from \secref{sec:pareto-frontier} hold for realistic code in a metaprogramming system. \cpcap{} dominates \llvm{} \ozero{}, and there are fewer cases that are beneficial to compile at a higher optimization level. While higher optimization levels used to make sense when the average 38\% speedup over \ozero{} could amortize a \num{9.2}\X{} average increase in compilation time, with \cp{} a smaller 24\% speedup must amortize an average \num{1286}\X{} increase in compilation time. For example, TPC-H Q5 compiles in \SI{0.25}{\second} with \llvm{} \othree{} in our compiler, and the resulting code performs 1.2\% better than \cp{}. To pay back the cost of the \num{1435}\X{} compilation time increase over \cp, the query would need to run for \SI{21}{\second}. On the TPC-H data set, however, the query finished execution in less than \SI{0.1}{\second}. Furthermore, in an industry-strength database, compilation would take several times longer, but execution would likely be faster due to better-engineered execution strategies, rendering the gap even larger.

\subsubsection*{Comparison to AST Interpretation}

The \cp{} algorithm outperforms our AST interpreter on the TPC-H benchmarks by an order of magnitude, as shown in \figref{fig:tpch-interp} (right). As shown in \figref{fig:tpch-interp} (left), the startup overhead of \cpshort{} is two--three times higher than the interpreter, but both are so small that they are negligible: in most cases it takes longer to construct the AST. For example, it takes \cp{} \SI{178}{\micro\second} to generate code for TPC-H Q5, but constructing the AST from query plan already takes \SI{326}{\micro\second}. And it takes the interpreter \SI{1.17}{\second} to execute it. Thus, \cp{} essentially completely replaces interpreters for database query execution.

\subsection{Copy-and-Patch Scalability}

As a baseline compiler, the \cp{} algorithm runs in linear time, requiring only two traversals of the AST and one traversal of the CPS call graph. Optimizing compilers like \llvm{}, on the other hand, contain non-linear algorithms. \figref{fig:eval-scaling} shows how the \llvm{} optimization levels scale as the input program size grows, on a synthetic function containing a sequence of statements that increment a variable by another variable. The performance of \llvm{} \ozero{} bogs down in instruction selection, while the higher optimization levels spend their time collapsing the increments into a single resulting statement. In both cases, however, \llvm{} compilation is increasingly slow compared to \cp{} as the source code size increases.

\subsection{Copy-and-Patch Optimization Breakdown}

\begin{figure}
    \begin{minipage}{0.45\linewidth}
        \centering
        \includegraphics[width=\linewidth]{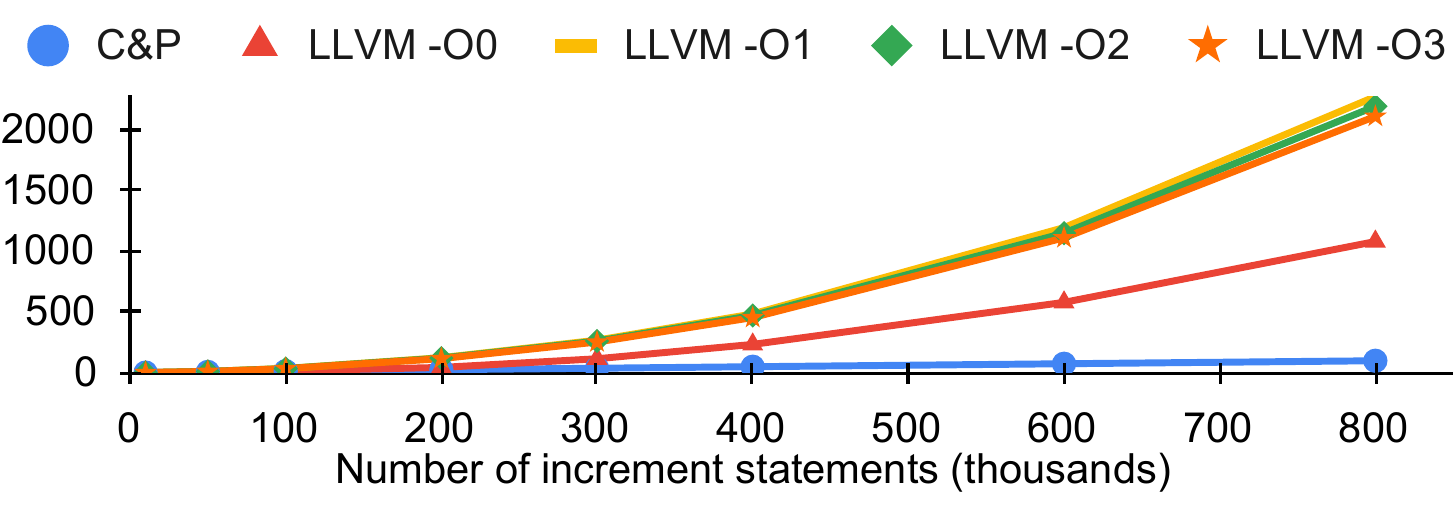}
        \vspace{-1.3em}
        \caption{
            The normalized startup delay of \cpshort{} and LLVM optimization levels as the size of the input program increases. To demonstrate scalability, the time it took each algorithm to compile 10k statements is normalized to 1, so perfect scaling line would end at y-axis 80. \cpshort{} scales near-linearly (ending at y-axis 98), while the other lines show worse scalability.
            \label{fig:eval-scaling}
            \vspace{-1em}
        }
    \end{minipage}
    \hfill
    \begin{minipage}{0.53\linewidth}
        \centering
        \vspace{-0.4em}
        \includegraphics[width=0.95\linewidth]{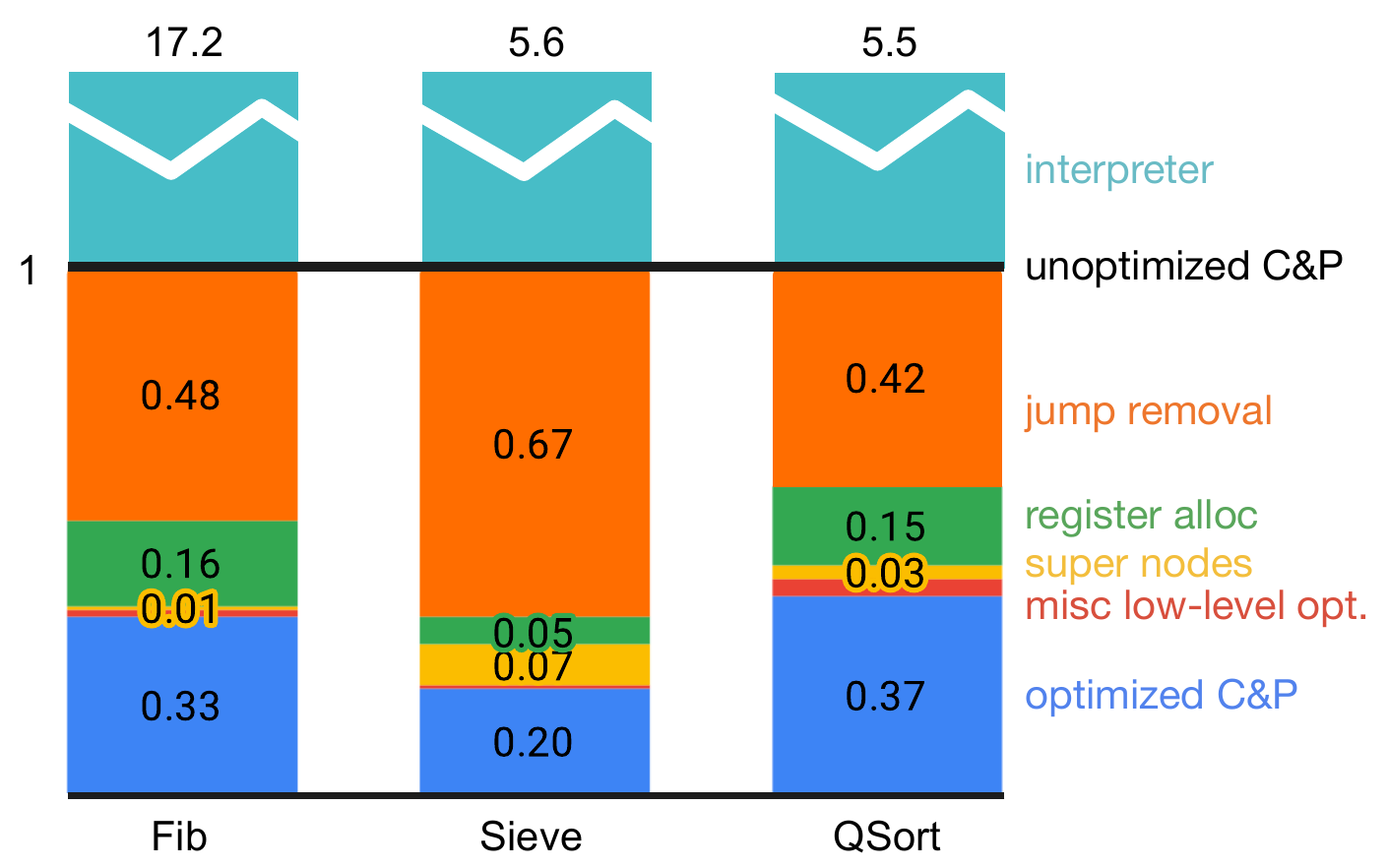}
        \vspace{-1em}
        \caption{
            Normalized microbenchmark running times with optimized \cpshort{} in dark blue. Each bar on top of that shows the running time added by removing each optimization.
            \label{fig:breakdown}
            \vspace{-1em}
        }
    \end{minipage}
\end{figure}

\cpcap{} employs several optimizations to produce fast code. \figref{fig:breakdown} shows the impact of these optimizations on the three microbenchmarks as a stacked bar graph. The runtimes are normalized so that unoptimized \cpshort{} is at one. The blue bars show the runtime of the optimized versions, and each bar stacked on top shows the runtime added by removing one optimization. The largest gain comes from the core of the \cpshort{} technique, which generates specialized AST node implementations with direct branch instructions and directly-embedded runtime constants, yielding a \num{5.5}\X{} to \num{17.2}\X{} speedup compared with an interpreter. Inside \cpshort{}, jump removal and light-weight register allocation accounts for the bulk of the runtime saved through optimization. The Sieve benchmark also benefits from the supernodes, because it is more dominated by memory accesses and arithmetic expressions than the other benchmarks. Finally, a few low-level optimizations, such as instruction block aligning, account for the last part of the optimization gains.  

\subsection{Other Costs and Metrics}

The AST in our implementation consists of \astNodeNum{} types of nodes that implement the statements and expressions of an imperative language with some object-oriented capabilities.

The stencil library consists of \stencilNum{} stencils and is constructed at library installation time in \stencilGenTime{} using \stencilGenThreads{}. The code and data in the stencil library consume \stencilSize{} of memory at runtime. In contrast, the LLVM library consumes \llvmSize{} of memory at runtime in our build. The numbers are measured by computing the total size of symbols belonging to the respective library (whose names start with {\tt llvm::} and {\tt stencil::} respectively), using the {\tt nm} Linux command.

Finally, adding an AST node requires adding a stencil generator, but the required effort is modest. For example, adding a new generator for an \texttt{add} operator with SQL overflow semantics takes 82 lines of code, 17 of which are new logic, while the rest are boilerplate copied from other generators.


\section{Related Work}
\label{sec:related-work}

We compare the \cp{} algorithm and our MetaVar system with work in three areas:  baseline compilers, staged compilation that generates binary code by symbolically executing an interpreter, and build-time code library generation approaches that contrast to MetaVar.

Depending on how the code is generated, baseline compilers can be categorized into two classes: baseline compilers that use platform-dependent machine instruction assemblers to generate code, and baseline compilers that generates code by concatenating pre-compiled binary snippets. The term ``template JITs'' is often used to refer to both, but for the purpose of distinguishing them, we will call the former ``assembler-based baseline compiler'' and only the latter ``template JITs''.

\subsection*{Template JITs}

Template JIT is the technique that is most similar to \cp{}. A template JIT uses an optimizing compiler (e.g., C, Java) to generate binary snippets at build time, and at runtime generate code by concatenating those binary snippets. Examples include~\cite{maxine2013, templateJit2003, piumarta1998, ertl2003}. The Maxine compiler~\cite{maxine2013} generates snippets from Java functions, while the other systems generate snippets from C functions. 

There are three major differences between \cp{} and the template JIT techniques described above. First, \cp{} has a patching phase. None of the papers above supported patching the binary code to burn in literals, stack offsets, and jump addresses, so their technique only works if the binary code can be concatenated without modification. This implies that all jumps and calls are indirect, and that all constants must be retrieved from memory, resulting in inferior execution performance. Second, \cp{} has the concept of stencil variants, which allows \cp{} to not only generate higher-quality code by selecting the most matching variant (e.g., adding a value in register with a constant), but also perform optimizations like register allocation and super-instructions, thus improving execution performance further. Third, all above papers target low-level bytecode assembling. In contrast, the use of a CPS call graph, and the flexibility of the binary stencils allow \cp{} to be used not only for low-level bytecode assembling, but also for high-level language compilation.

\subsection*{Assembler-Based Baseline Compilers}
\label{sec:assemblers}

Using a machine code assembler to generate code is another approach to baseline compilers. The baseline compiler decide what assembly instruction to emit, and then the machine code assembler assembles it to machine code. 
Baseline compilers like the Google Chrome Liftoff~\cite{liftoffBlog}, the Firefox WebAssembly baseline compiler~\cite{firefoxWasmBaselineBlog1}, and Wasmer SinglePass~\cite{wasmerSinglepassJit} all employ this approach. Unsurprisingly, a lot of work has been done to improve the performance of machine code assemblers. VCODE~\cite{engler1996} proposed to use a library of hand-written platform-specific instruction implementations to speed up the assembling process, and this approach was also followed in AsmJIT~\cite{asmjit}, DynASM for Lua~\cite{dynasm}, and as a case study in Terra~\cite{devito2014}. The DCG system~\cite{engler1994} also attempts greedy register allocation, but only works under the unrealistic assumption that no spilling is ever needed, and also runs 35 times slower than VCODE~\cite{engler1996}. 

In the world of bytecode assemblers, \cp{} has two advantages compared with baseline compilers using a machine code assembler. First, as shown in \secref{sec:webassembly}, \cp{} not only emits code faster, but also emits faster code.
Second, since the stencils are generated by Clang, we don't need to figure out what assembly instruction to use, thus reducing the engineering cost.

In the world of full compilers, the benefit of our approach is reflected in both engineering cost and performance. When the task is to lower a high-level program to machine code, assembling the instructions is not the most difficult part. Deciding \textit{what} to assemble is. We need optimization, assembly instruction selection, and register allocation to produce high quality code. A machine code assembler cannot provide any of these. \cpcap{} solves the problem by using Clang to generate the AST node stencils. By offloading all low-level and architecture-specific details to Clang, we avoid the prohibitively high engineering cost of re-inventing the big wheel of target-optimized instruction selection and assembling for every architecture,
and is portable to any architecture supported by \llvm{}. Furthermore, \cp{} pushes the CPU cost of register allocation, instruction selection and assembling to library build time. At runtime, it only copies pre-built chunks of instructions, which is clearly faster than doing register allocation and then selecting and assembling each machine instruction. 


\subsection*{Techniques Originated in Other Areas}

Continuation-passing style~\cite{steele1977}, which is similar to threaded code~\cite{bell1973}, is originally used to optimize an interpreter's performance, as well as in compilation of functional languages. We use this technique to weave together the control flow of stencils. In this technique, control flow passes through an AST bottom-up, letting us convert calls to jumps. 

Superinstruction~\cite{proebsting1995, javaSuperInstruction} is a well-known technique to reduce indirect jump overhead between interpreter opcodes. Our supernode is similar to superinstruction; but in our use case, since unnecessary jumps between opcodes are already eliminated, supernodes are only a modest optimization to improve the quality of generated code.

The idea of using external variables to locate holes to burn in runtime constants is used by~\citet{noel1998}. However, their use case is runtime specialization of statically known logic and cannot be used in our use case where the logic is generated at runtime. Their technique is also more verbose than ours. QEMU~\cite{qemuPaper} also used this trick to translate CPU instructions between architectures, but requires non-standard GCC extensions that has since been removed.

\subsection*{Staged Compilation and Dynamic Specialization}

Staged compilation~\cite{thibault2000,consel1998} and dynamic specialization~\cite{finkel2019}, given an interpreter implementation and an opcode sequence, generates specialized optimized binary code by symbolically evaluating the interpreter on the opcode sequence. The major advantage is that the user only needs to write an interpreter backend, and the specializer automatically generates binary code. This code generation process, however, runs slower than a hand-written backend, so it is not suitable for the use case where compilation time matters.

\subsection*{Build-time Code Library Generation}

Like the stencil library, FFTW~\cite{frigo1998} employs a code library approach. At compilation time, it creates a collection of \textit{codelets} that implement optimized variants of FFT on various fixed-length input sequences. It has a dedicated compiler that generates optimized C code implementing codelets. At runtime, input is split into smaller pieces using a divide-and-conquer strategy and, when a piece is small enough, it is dispatched to one of the pre-built codelets. The FFTW codelet library is similar in spirit to the stencil library, except that it only contains implementations of FFTW and it does not burn in constants.  The FFTW algorithm calls codelets by indirect function calls,  which is acceptable because each pre-built implementation does significant work. Nevertheless, the paper mentioned that reading runtime constants from memory hurts performance. We envision that our MetaVar system's ability to burn runtime constants into instruction flow and make indirect jumps and calls direct could further improve the performance of the generated FFTW codelets.

\section{Conclusion}
\label{sec:conclusion}

We introduced \cp{}, a novel compilation technique that generates decent executable code at a negligible compilation cost. We envision \cp{} used in domains where fast compilation is important, notably, as the baseline compilation tier in JIT compilers. We empirically evaluated its potential in two such domains: SQL query compilation and WebAssembly compilation. In both domains, we demonstrated that \cp{} significantly outperforms existing techniques and approaches for fast compilation, including the current state-of-the-art baseline compiler implementations (V8 Liftoff and Wasmer Singlepass), \llvm{} \ozero{} compilation, and interpreters. 

The \cp{} algorithm and the stencil generator system are extensible by design. New AST nodes can be added by users seeking better performance for a new, perhaps domain-specific, language construct. And new supernode stencil generators can be added for better local optimization. The type system of the AST can also be expanded in future work to include vector types to target vectorized instructions. Therefore, we believe \cp{} will find a place in domains other than SQL query compilation and WebAssembly compilation as well.

We envision two areas of future work. First, new general-purposed optimizations can be implemented, such as common subexpression elimination, loop unrolling, and vectorization. Of course, these will increase compilation time, but will do so starting from a lower starting point. Second, \cp{} can be combined with domain-specific optimization techniques, most notably, the type profiling, type speculation, and inline caching techniques that optimize the performance of dynamic-typed languages. With these optimizations, we believe \cp{} can also be used as a fast profiling tier or even a fast optimizing tier in dynamic language JIT engines.

\begin{acks}
We thank our anonymous reviewers for their comments that helped us improve this manuscript. We would also like to thank
Alex Aiken,
Saman Amarasinghe,
Saam Barati, 
Ajay Brahmakshatriya,  
Cheng Chen,
Stephen Chou,
David Durst,
Slava Egorov,
Lang Hames,
Pat Hanrahan,
Scott Kovach,
Richard Peng,
Zhou Sun, 
Leszek Swirski, and
Yinzhan Xu
for helpful comments, review, and references. This work was supported by the Stanford Agile Hardware Center.
\end{acks}

\bibliography{paper}

\end{document}